\documentclass{article}

\usepackage{arxiv}

\usepackage[utf8]{inputenc} 
\usepackage[T1]{fontenc}    
\usepackage{hyperref}       
\usepackage{url}            
\usepackage{booktabs}       
\usepackage{amsfonts}       
\usepackage{nicefrac}       
\usepackage{microtype}      
\usepackage{lipsum}		
\usepackage{graphicx}
\usepackage{doi}

\title{Influence of plasma instabilities on the propagation of electromagnetic cascades from distant blazars}

\author{Luis E. Espinosa Castro \\
    Gran Sasso Science Institute, \\ 
    Viale Francesco Crispi 7, 67100, L’Aquila, Italy \\
    INFN, Laboratori Nazionali del Gran Sasso, \\ 
    Via G. Acitelli 22, 67100, Assergi, Italy \\
	\texttt{luis.espinosacastro@gssi.it} \\
	\And
	Simone Rossoni \\
    II. Institute for Theoretical Physics \\
    Hamburg University, \\ 
    Luruper Chaussee 149, 22761, Hamburg, Germany \\
	\texttt{simone.rossoni@desy.de} \\
    \And
	Günter Sigl \\
    II. Institute for Theoretical Physics \\
    Hamburg University, \\ 
    Luruper Chaussee 149, 22761, Hamburg, Germany \\
	\texttt{guenter.sigl@desy.de} \\
}

\renewcommand{\shorttitle}{}

\hypersetup{
pdftitle={A template for the arxiv style},
pdfsubject={q-bio.NC, q-bio.QM},
pdfauthor={David S.~Hippocampus, Elias D.~Striatum},
pdfkeywords={First keyword, Second keyword, More},
}

\begin{document}
\maketitle

\begin{abstract}
    The propagation of very-high-energy gamma-rays (VHEGRs) in the extragalactic space offers the opportunity to study astrophysical phenomena not reproducible in laboratories. In particular, the deviation from predictions of the observed photon flux from distant sources at the GeV energy scale still represents an open problem. Commonly, this deviation is interpreted as the result of the deflection out of the line of sight of the source of the cascade-produced charged leptons by weak magnetic fields present in the intergalactic medium (IGM). However, plasma instabilities could have an effect on the energy- and momentum-distribution of the secondary electrons and positrons, modifying the development of electromagnetic cascades. In this work, we study the influence of plasma instabilities on the energy spectrum of electromagnetic cascades through a parametric study, performed with the Monte Carlo code CRPropa. We parameterize plasma instabilities  with an energy loss length normalisation $\lambda_0$ and a power law index $\alpha$ of its electron/positron energy dependence. We simulate photon spectra at Earth for different blazar scenarios and find that plasma instabilities can reproduce the suppression in the GeV-photon flux of real astrophysical sources, such as 1ES 0229+200, when the energy loss length of electrons/positrons at 1.0 TeV is $\lambda_0 \simeq 100\,\text{kpc}$ and $\alpha\simeq0$. The energy fraction lost by the secondary electron pairs due to the instability is estimated to be about $1\%$ over the typical interaction length of Inverse Compton scattering for these parameter values.
\end{abstract}

\keywords{Astroparticle physics \and blazars \and gamma-rays \and  electromagnetic cascades \and intergalactic magnetic field \and plasma instabilities}

\section{Introduction}
\label{sec_intro}
Extragalactic VHEGRs can originate from jet emission of active galactic nuclei (AGNs) or in prompt events, such as gamma-ray bursts (GRBs). These high energy photons propagate through extragalactic space, and they can interact with the photons of background cosmic radiation fields, such as the Cosmic Microwave Background (CMB) \cite{Dicke1965,Penzias1965} or the Extragalactic Background Light (EBL) \cite{Hauser2001}. These interactions result in the production of pairs of electrons and positrons (electromagnetic pair production (PP), $\gamma+\gamma_{\text{bg}}\rightarrow e^+ + e^-$, where $\gamma_\text{bg}$ is the backgroud photon). The produced charged leptons can subsequently interact with the background fields by inverse Compton scattering (IC, $e^\pm + \gamma_\text{bg} \rightarrow e^\pm + \gamma$). These processes continue to repeat until the energy of the secondary photons is below the threshold to produce more electro-positron pairs, defining the so-called electromagnetic cascade. After this point, the remaining lepton pairs continue to undergo inverse Compton scattering, populating the photon flux at the GeV scale. In addition to these interactions, higher order processes, like double (DPP, $\gamma+\gamma_\text{bg}\rightarrow e^+ + e^- + e^+ + e^-$) and triplet (TPP, $e^\pm+\gamma_\text{bg}\rightarrow e^\pm + e^+ + e^-$) pair production are also possible. However, these interactions are very suppressed below $\sim 100\,\text{TeV}$ (for more details on interaction rates see \cite{Heiter:2017cev}), and are characterized by a higher interaction threshold. Electromagnetic cascades act as a calorimeter, in the sense that the total energy injected above the PP energy threshold will roughly equal the energy of the produced cascade. Fully developed cascades have been shown to have a universal energy spectrum, independent of the photon spectrum of the source and its distance \cite{Berezinsky2016}. The propagated spectrum is described by a power-law $J(E)\propto E^{-\Gamma}$ with $\Gamma=1.9$ for the cascade multiplication stage (and $\Gamma=1.5$ from IC cooling at even lower energies). Based on this consideration, for astrophysical sources characterized by a photon spectrum harder than $\Gamma=1.9$ the photon flux below a few hundred GeV on Earth will be dominated by cascade photons. On the contrary, sources with softer spectrum at injection will show a propagated GeV spectrum dominated by prompt photons.
\par Observation of VHEGRs provides information not only on the sources in which they are produced, but also on the medium in which they propagate. Measurements of the photon flux coming from electromagnetic cascades can be performed using ground- and space-based detectors. The first class corresponds to detector arrays that register the extensive air showers produced by incoming photons when they interact with the Earth's atmosphere. Some examples include the Imaging Atmospheric Cherenkov Telescopes such as H.E.S.S. \cite{Krawczynski1999}, MAGIC \cite{Saggion2006} and VERITAS \cite{Holder2006}. The case of space-based detectors consists of space missions for direct detection of gamma-rays. Among them, the Fermi Gamma-ray Space Telescope and its Large Area Telescope (Fermi-LAT) \cite{Atwood2009} is an example.
\par A prominent problem in the understanding of VHEGRs from distant sources is the discrepancy between the predicted and observed photon flux at the GeV energy scale. The observed photon spectrum shows a suppression in the GeV range compared with the expected spectrum. One possible explanation is the magnetic deflection of the secondary electron-positron pairs by weak intergalactic magnetic field (IGMF). This weak field, with lower bounds of the order of $10^{-15}\,\text{G}$ to $10^{-18}\,\text{G}$ (see \cite{Acciari2022,Vovk2023}), could induce a broadening of the momentum distribution of the electrically charged component of the cascade, beyond the detector's field of view. In addition, the IGMF could also introduce a delay in the arrival time of the secondary photons produced by the deflected charged leptons. So far, it is not possible to measure the IGMF directly. However, several studies have imposed constraints on the root mean square value of the magnetic strength $B_\text{rms}$ and the coherence length $\lambda_\text{c}$, from the observation of the photon spectrum of astrophysical sources \cite{DaVela2023,Korochkin2021,Neronov2009, Neronov2010, Taylor2011}. This is based on simulating the magnetic deflection of electron pairs during the cascade development using different IGMF configurations, and comparing the propagated spectrum with observations from ground and space-based detectors.
\par Another possible explanation for the observed suppression can be given by the effect of plasma instabilities on the charged component of the electromagnetic cascade. Plasma instabilities are the result of the excitation of unstable electrostatic waves generated by an electron-positron beam, giving rise to the growth of small perturbations in the intergalactic medium (IGM). These plasma oscillations are induced by the change in the electron density resulting from the electron bunching and subsequent repulsion \cite{Tonks1929}. The influence of the instabilities on the cascades is given by the resonant interaction between the electron-positron pairs and the plasma waves, which influences the energy- and momentum-distribution of the particles. In general, plasma instabilities can have two main effects: the angular broadening of the cascade, and the energy drain from the pairs to heat the IGM \cite{Alawashra2024,Perry2021}. 
\par In recent years, plasma instabilities have gained attention in astroparticle physics. Previous studies \cite{Perry2021, Alawashra2022,AlvesBatista2019} were focused on the efficiency with which plasma instabilities can grow in the IGM. Their impact was evaluated for different initial conditions, such as the angular aperture of the injected beam, the instability growth rates and the state of the IGM, as defined by its temperature and number density. In these studies, it was obtained that the induced energy loss of the charged component of the cascade may reproduce the flux suppression related to the IGMF for certain AGN emissions, but may be small for initially very-narrow beams, propagating in weak IGMF. In this work, we consider the electron-positron energy loss term associated with plasma instabilities, and perform a study of a general \textit{two-variables} parameterization of the typical energy loss length of instabilities. In particular, we implement this process in a modern simulation code for gamma-ray propagation.
\par CRPropa 3\footnote{More information about the code can be found at \href{https://crpropa.github.io/CRPropa3/index.html}{https://crpropa.github.io/CRPropa3/index.html}} \cite{AlvesBatista2016} is a public Monte Carlo code for the simulation of the extragalactic propagation of ultra-high-energy cosmic rays (UHECRs), electromagnetic particles and neutrinos. In this work, the development of the electromagnetic cascades initiated by a steady source of VHEGRs is simulated with the code version CRPropa 3.2 \cite{AlvesBatista2022}. Thanks to the modular structure of the code, new physical processes can easily be included in the main code, via a new simulation module. We simulate the role of the energy drain of plasma instabilities on electron-positron pairs with the implementation of a newly developed module. Therefore, we evaluate the dependence of simulation results on the variation of the parameters of our instability model. We also consider the variation of flux suppression with respect to the parameters of the photon spectrum at the source. 
\par This paper is organised as follows. The  modeling of the energy loss term is described in Section \hyperlink{Model}{2}. The setup of our simulations and analysis of the results are described in Section \hyperlink{Cascadesimulations}{3}. The results obtained for different instability parameters and injection scenarios are shown and described in Section \hyperlink{Results}{4}. Lastly, in Sections \hyperlink{Discussion and conclusions}{5}, the interpretation of these results and conclusions are discussed.

\hypertarget{Model}{}
\section{Model}
\label{sec_model}
The temporal evolution of the momentum distribution function $f(\textbf{p},t)$ (i.e. the number of particles per unit of momentum and unit of time) is given by the Fokker-Planck (FP) equation. The FP equation is a partial differential equation in time and particle momentum, and it reads
\begin{equation}
\label{eq:2.1}
    \dfrac{ \partial f(\textbf{p},t)}{\partial t} = - \sum_{i=1}^3 \dfrac{ \partial }{\partial p_i} \left[V_i(\textbf{p},t)f(\textbf{p},t)\right] + \sum_{i,j=1}^3 \dfrac{ \partial }{\partial p_i} \left[D_{ij}(\textbf{p},t)\dfrac{ \partial f(\textbf{p},t)}{\partial p_j}\right]\,,
\end{equation}
where $\textbf{p}$ is the momentum vector with components $\textbf{p}=(p_1,p_2,p_3)$, $V_i(\textbf{p},t)$ and $D_{ij}(\textbf{p},t)$ are the drift-vector and diffusion-tensor coefficients, respectively, as defined in \cite{Nicholson1983}. The temporal evolution of the momentum distribution function is given by the interplay of two terms. The first term on the right hand side of \eqref{eq:2.1} corresponds to the energy drain due to overall momentum losses, while the second term is related to the angular and longitudinal widening caused by momentum diffusion. 
\par We formulate a simplified model of the energy losses induced by plasma instabilities on the electron-positron pairs. Considering only the analysis of the beam energy changes in the kinetic regime of the instabilities \cite{Perry2021}, the second term on the right hand side can be neglected. This simplification is further justified by numerical studies of pair-beam induced plasma instabilities \cite{Alawashra2024, Perry2021}, which suggest very small angular broadening caused by plasma instabilities, of the order of $\sim10^{-4}$ radians. Therefore, in the special case of a one-dimensional stationary source, the drift coefficient can be assumed to be independent of the particle momentum, to further simplify the FP equation, as discussed in \cite{Beck2023}. 
\par In order to perform a parametric study of this scenario, we develop a numerical energy-loss module associated to plasma instabilities. It has been shown in \cite{AlvesBatista2019} that the instability growth time $\tau_\text{PI}$ can be modeled as a power-law of the form \(\tau_\text{PI} \propto E_e^\alpha\), where $E_e$ is the electron energy and $\alpha$ is the instability power index. In this work, we consider instability power index values such that $-2\leq \alpha \leq 2$, without referring to any specific instability model\footnote{The rage of values of the index $\alpha$ was chosen based on the electron energy dependence of the instability growth time discussed in the plasma instability models in \cite{AlvesBatista2019}}. In \cite{AlvesBatista2019}, the effect of the instability is maximized to find a robust lower limit of the suppressed flux by considering the energy-loss time scale of the electron-positron pairs to correspond to the instability growth time, although in principle the energy-loss time can become larger. Therefore, the typical length scale of plasma instabilities $\lambda_{PI}$ in the ultra-relativistic limit can be modeled as \(\lambda_{PI} \propto c\cdot E_e^\alpha\), where $c$ is the speed of light. Then, the energy-loss-per-unit-length can be written as   
\begin{equation}
\label{eq:2.2}
    -\dfrac{dE_e}{dx} =\dfrac{E_e}{\lambda_\text{PI}} \propto E_e^{1-\alpha}\,.
\end{equation}
In order to normalize the energy-loss-per-unit-length in Equation \eqref{eq:2.2}, we compare the value of the plasma instability length scale $\lambda_\text{PI}$ with IC interaction length. We consider the electron energy value \(\widetilde{E}_e\) at which the length scale of plasma instabilities and the IC interaction length are comparable. This reads
\begin{equation}
\label{eq:2.3}
    \lambda_\text{PI}(\widetilde{E}_e)=\lambda_\text{IC}(\widetilde{E}_e)\,,
\end{equation}
where $\lambda_\text{IC}(\widetilde{E}_e)$ is the IC interaction length for electrons with energy $\widetilde{E}_e$. Consequently, if we define the IC length scale at $\widetilde{E}_e$ with $\lambda_0=\lambda_\text{IC}(\widetilde{E}_e)$, the typical length scale of the instabilities is given by the following relation
\begin{equation}
\label{eq:2.4}
    \lambda_{PI}(E_e)=\lambda_0\left(\frac{E_e}{\widetilde{E}_e}\right)^\alpha\,.
\end{equation}
Using Equations \ref{eq:2.2} and \ref{eq:2.4}, we can write the energy-loss-per-unit-length as 
\begin{equation}
\label{eq:2.5}
    -\frac{dE_e}{dx} = \frac{\widetilde{E}_e}{\lambda_0}\left(\frac{E_e}{\widetilde{E}_e}\right)^{1-\alpha}\,.
\end{equation}
\par From both Equations \eqref{eq:2.4} and \eqref{eq:2.5}, we can see that the energy-loss term is defined by three parameters: the instability power index $\alpha$, the normalization electron energy $\widetilde{E}_e$ and the length scale $\lambda_0$. However, these three parameters are not independent. Indeed, the parameter combinations given by $\lambda_0$ and $\alpha$ can be always reproduced by the corresponding combinations of $\widetilde{E}_e$ and $\alpha$. For this reason, we decide to perform our parametric study by varying only $\alpha$ and $\lambda_0$. Results for different values of $\widetilde{E}_e$ can be obtained by calculating the corresponding value of the length scale $\lambda_0$.
\par In this work, we assume the normalization energy \(\widetilde{E}_e=10^{12}\) eV. At this energy the IC interaction length with the CMB can be approximated as in \cite{Neronov2009} with $\lambda_\text{IC}=1.2$ kpc at zero redshift. Hence, we set $\lambda_0=1.2$ kpc as one of the possible values for the length scale of the plasma instabilities. To investigate the impact of plasma instabilities on the observed photon flux, we consider several values for the parameters $\alpha$ and $\lambda_0$. In particular, we consider $\alpha = -2.0,\,-1.5,\,...,\,1.5,\,2.0$ with $\Delta\alpha=0.5$, and $\lambda_0 =0.12,\,1.2,\,12.0,\,120.0\,\text{kpc}$. 
\begin{figure}[t]
\centering
\begin{minipage}{7.65cm}
\centering
\includegraphics[scale=0.45]{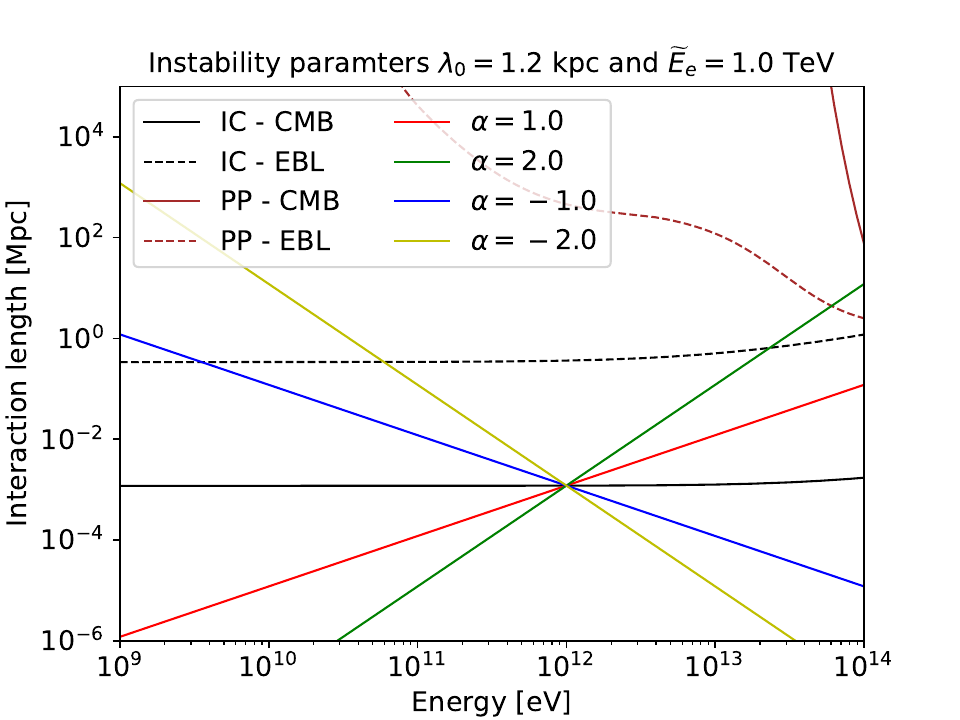}
\end{minipage}
\begin{minipage}{7.65cm}
\centering
\includegraphics[scale=0.45]{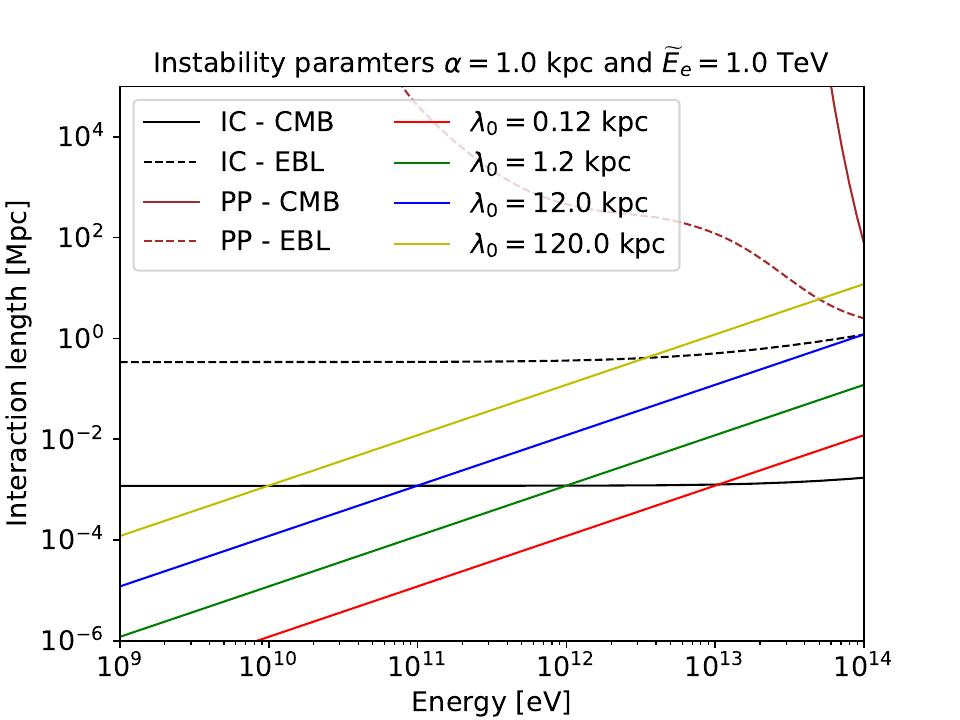}
\end{minipage}
\caption{Inverse Compton scattering (IC, black lines) and pair production (PP, brown lines) interaction lengths. Solid lines refer to interactions with CMB photons and dashed lines with EBL (from \cite{Franceschini2008}). LEFT: plasma instability typical length scale $\lambda_0=1.2\,\text{kpc}$ and  different values of the instability power index $\alpha$. RIGHT: instability power index $\alpha=1.0$ and different values of the typical length scales $\lambda_0$. The normalization energy is $\widetilde{E}_e=1.0\,\text{TeV}$ in both the panels.}
\label{fig:fig1}
\end{figure}
\par In Figure \ref{fig:fig1} the interaction lengths in \eqref{eq:2.4} are shown as a function of energy, for several plasma instability parameters considered in this work. The left and right panels correspond to different values of the index $\alpha$ and the length scale $\lambda_0$, respectively. Equation \eqref{eq:2.4} is shown along with the IC scattering and pair production interaction lengths, for both the CMB and EBL photon fields (the EBL model used in this study is the one from \cite{Franceschini2008}). From Figure \ref{fig:fig1} we can deduce some qualitative behaviour of electron and positron interactions (in this section we will refer only to electrons, but the same results apply to positrons). In the left panel of Figure \ref{fig:fig1} we can see that for $\lambda_0=1.2\,\text{kpc}$, positive  plasma indexes ($\alpha>0$) imply that electrons with energy $E_e>1\,\text{TeV}$ will preferentially undergo IC scattering with the CMB photons producing secondary photons. On the contrary, for electrons with $E_e<1\,\text{TeV}$ plasma cooling is more efficient than IC scattering. In this case, the electron energy will be reduced without producing secondary photons contributing to the cascade. The exact opposite behaviour is expected when the plasma index is $\alpha<0$. 
\par In the left panel of Figure \ref{fig:fig1} it can be seen that the plasma energy loss length and the IC length intersect at $E_e=1\,\text{TeV}$ (i.e. at the electron normalization energy chosen). The Thomson regime is given by the condition $s-m_e^2 \ll m_e^2$, where $s$ is the square of the center of mass energy and $m_e$ is the electron mass (for more details see \cite{Jackson1998,Sigl2017}). This condition corresponds to the following relation for the electron energy
\begin{equation}
\label{eq:2.6}
E_e\ll\dfrac{m_e^2}{4\epsilon}\,,
\end{equation}
where $\epsilon\sim k_B T_0$, with $k_B$ the Boltzmann constant and $T_0$ the CMB temperature, is the photon energy in the CMB frame. Numerically, this condition corresponds to $E_e \lesssim 10^3 \,\text{TeV}$. In this regime, the energy of the IC up-scattered photon is
\begin{equation}
\label{eq:2.7}
\epsilon_\text{up} \simeq \dfrac{4\epsilon E_e^2}{3m_e^2} = 1.5\cdot\left( \dfrac{\epsilon}{1\,\text{meV}}\right)\left( \dfrac{E_e}{10\,\text{TeV}}\right)^2 \text{TeV}\,.
\end{equation}
In our scenario, with $E_e=1\,\text{TeV}$ and $\epsilon\sim k_B T\sim 0.7\,\text{meV}$ \cite{Fixsen2009}, we have $\epsilon_\text{up}\sim 10\,\text{GeV}$. Therefore, for $\lambda_0=1.2\,\text{kpc}$ we expect the transition from the IC-dominated cascade to the plasma instability-dominated cascade to occur around $10\,\text{GeV}$ in the propagated photon energy spectra. In the case of a different value of $\lambda_0$ (right panel of Figure \ref{fig:fig1}), the corresponding transition energy in the propagated photon spectrum can again be obtained from equation \eqref{eq:2.7} by using the electron energy $E_e$ for which $\lambda_\text{PI}=\lambda_\text{IC}$.

\hypertarget{Cascadesimulations}{}
\section{Cascade simulations}
\label{sec_simulations}
The propagation of VHEGRs in the extragalactic space requires to take into account all relevant interactions and energy loss processes for photons, electrons and positrons. Instead of directly simulating plasma instabilities in an electron-positron pair beam, as done in other studies such as  \cite{Alawashra2024}, we parameterise the energy-loss-per-unit-length due to the instabilities and simulate the development of the electromagnetic cascade initiated by VHEGRs.
\par For the numerical simulation of the electromagnetic cascades, we used the Monte Carlo code CRPropa 3.2. We take into account pair production, inverse Compton scattering, as well as double and triplet pair production. As background photon fields we consider the CMB and EBL, and for the latter we consider the model in \cite{Franceschini2008}. We include the new module for the instability energy loss term, described in Section \ref{sec_model}, in the same numerical framework. 
\par We consider the emission of a source characterized by a stationary jet emitting high-energy gamma-rays. In particular, the injected energy spectrum follows a power-law distribution $J_0(E_0)\propto E_0^{-\gamma}$, where $\gamma=1.0$ is the spectral index and $1\,\text{GeV}\leq E_0 \leq 100\,\text{TeV}$. We simulate the propagation of $10^5$ primary photons, with a propagation step size of $10^{14}$ cm (i.e. $10^{-4}\,\text{yr}$). The distance to the detection point is $L=580$ Mpc (i.e. the approximate distance of the blazar 1ES 0229+200 from \cite{Taylor2011}, which is taken as the reference scenario in our study). Since no deflection processes are included in our study, we simulate the propagation of the injected and created particles in one dimension.

\hypertarget{Outputanalysis}{}
\subsection{Output analysis}
\label{subsec_output_analysis}
As stated above, we simulate the propagation of VHEGRs injected with energy spectrum $J_0(E_0)\propto E_0^{-\gamma}$. In order to study different blazar emission scenarios, we reweigh our simulations to a more general spectral function at the injection. We computed the cascade signal function $G(E_0,E)$, where $E_0$ is the photon energy at the injection, and $E$ is the photon energy at the observation. As the number of photons increase during the development of the cascade, the photon observed with energy $E$ will not correspond to the same photon injected with energy $E_0$, but to one of the photons produced by the interactions. The signal function $G(E_0,E)$ is a two-dimensional function which contains the probability that a photons detected with energy $E$ was produced by a photon with initial energy $E_0$. The photon spectrum at Earth $J(E)$ can be then reconstructed from the convolution of the cascade signal and the injection spectrum $J(E_0)$. This reads
\begin{equation}
\label{eq:3.2}
    J(E) = \int_{E_0 \geq E} G(E_0,E)\cdot J(E_0)\,dE_0\,,
\end{equation}
as described in \cite{Acciari2022}. We assume an injected energy spectrum described by a generic power-law function times an exponential cutoff of the form
\begin{equation}
\label{eq:3.3}
    J(E_0) = J_0\cdot \left(\frac{E_0}{1\,\text{TeV}}\right)^{-\beta}\exp\left(-\dfrac{E_0}{E_{\text{cut}}}\right)\,,
\end{equation}
where $\beta$ is the source spectral index, $E_\text{cut}$ is the high energy cutoff and $J_0$ is a normalization constant. We will vary these parameters to study the effect of plasma instabilities on different injection scenarios. 
\par The simulated photon spectra can be compared to observed data by gamma-ray telescopes, such as Fermi-LAT. A statistical criterion is used to identify the parameters of our model of energy-loss-per-unit-length for plasma instabilities that agree better with the data. We calculate the $\chi^2$ in the following way

\begin{equation}
\label{eq:chi}
    \chi^2(\lambda_0,\alpha) = \sum_i \left(\frac{J_{\text{data}}(E_i)-J_{\text{sim}}(\lambda_0,\alpha,E_i)}{\sigma_{\text{data},i}}\right)^2\,,
\end{equation}
where $J_{\text{data}}(E_i)$ and $J_{\text{sim}}(\lambda_0,\alpha,E_i)$ are the values of the observed and propagated spectra at Earth for the same energy, respectively, and the sum is performed over the observed energy bins. The uncertainty for each value of the observed data is denoted as $\sigma_{\text{data},i}$. In this way, the combination of plasma instability parameters that best reproduce the observed photon spectrum will minimize $\chi^2$.

\hypertarget{Results}{}
\section{Results}
\label{sec_results}
In the first part of this section, we show the effects of varying the parameters $\alpha$ and $\lambda_0$ of our instability model on the propagated photon flux from a single blazar. In the second section, results corresponding to the change in the parameters of the injected energy spectrum of the blazar are discussed, specifically the spectral index $\beta$ and the cutoff energy $E_{\text{cut}}$. The effect of a different source distance is also considered.

\hypertarget{ResultsParametric}{}
\subsection{Dependence on instability parameters}
\label{subsec_results_parametric}
\begin{figure}[t]
\centering
\begin{minipage}{7.5cm}
\centering
\includegraphics[scale=0.3]{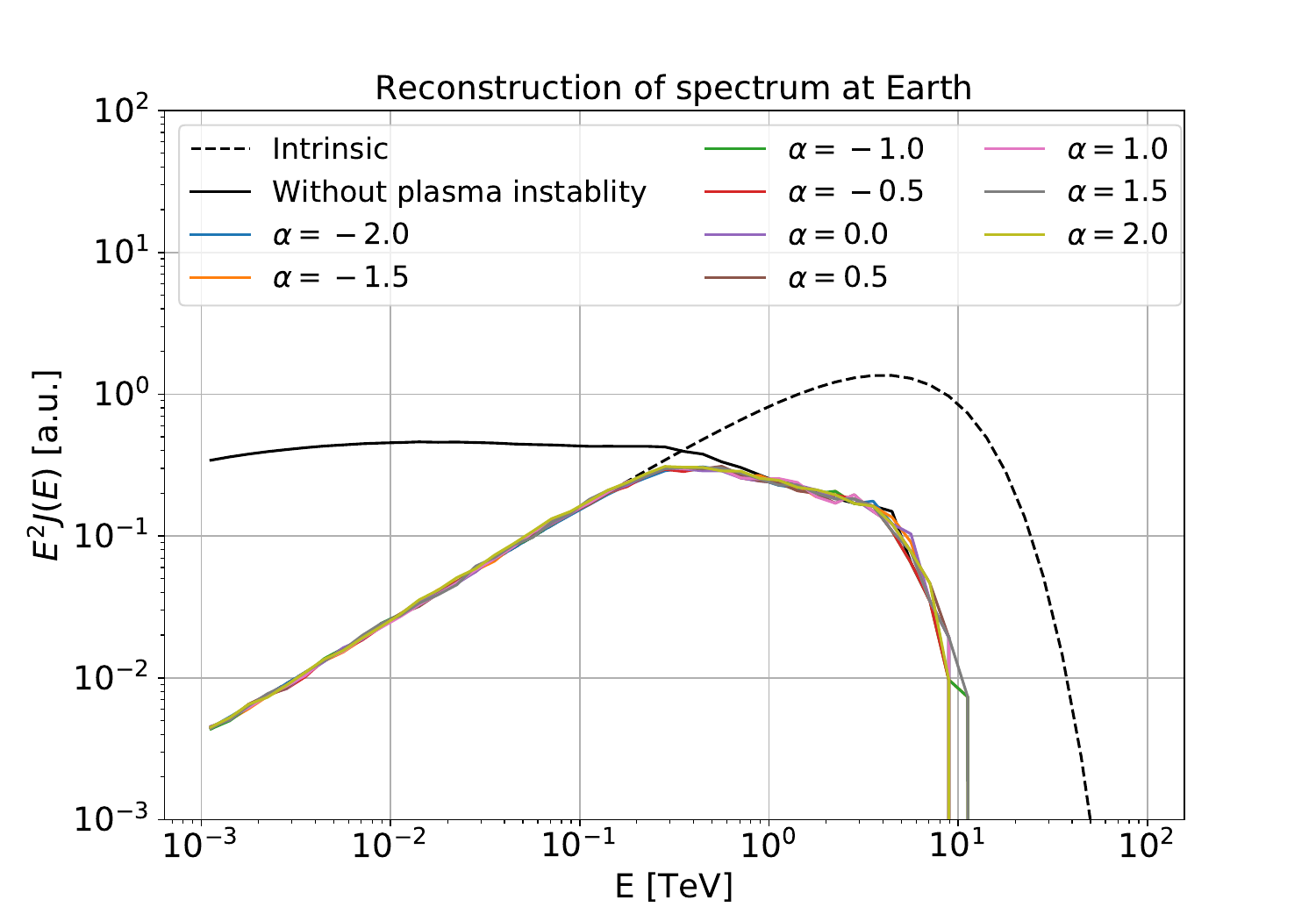}
\end{minipage}
\begin{minipage}{7.5cm}
\centering
\includegraphics[scale=0.3]{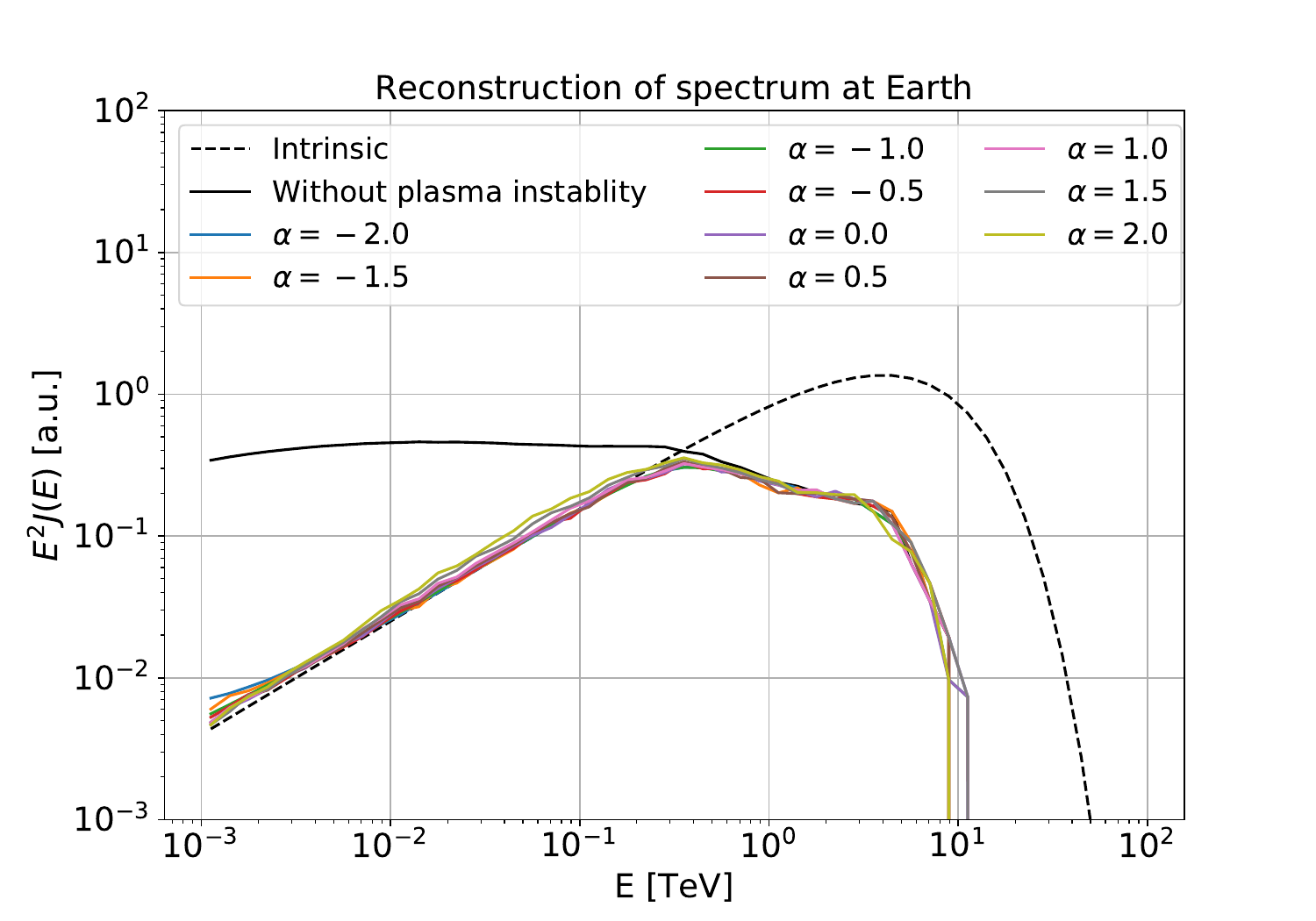}
\end{minipage}
\begin{minipage}{7.5cm}
\centering
\includegraphics[scale=0.3]{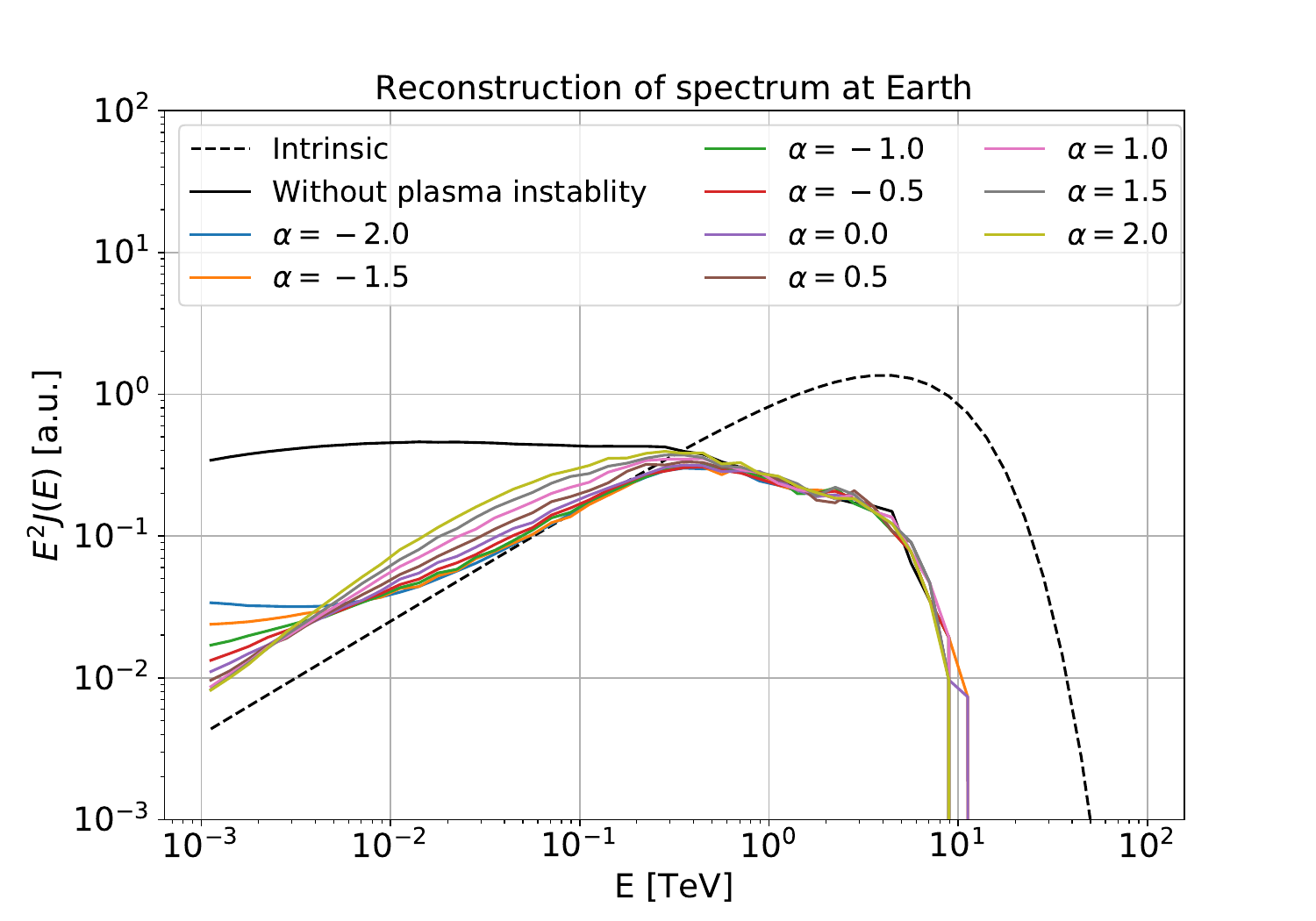}
\end{minipage}
\begin{minipage}{7.5cm}
\centering
\includegraphics[scale=0.3]{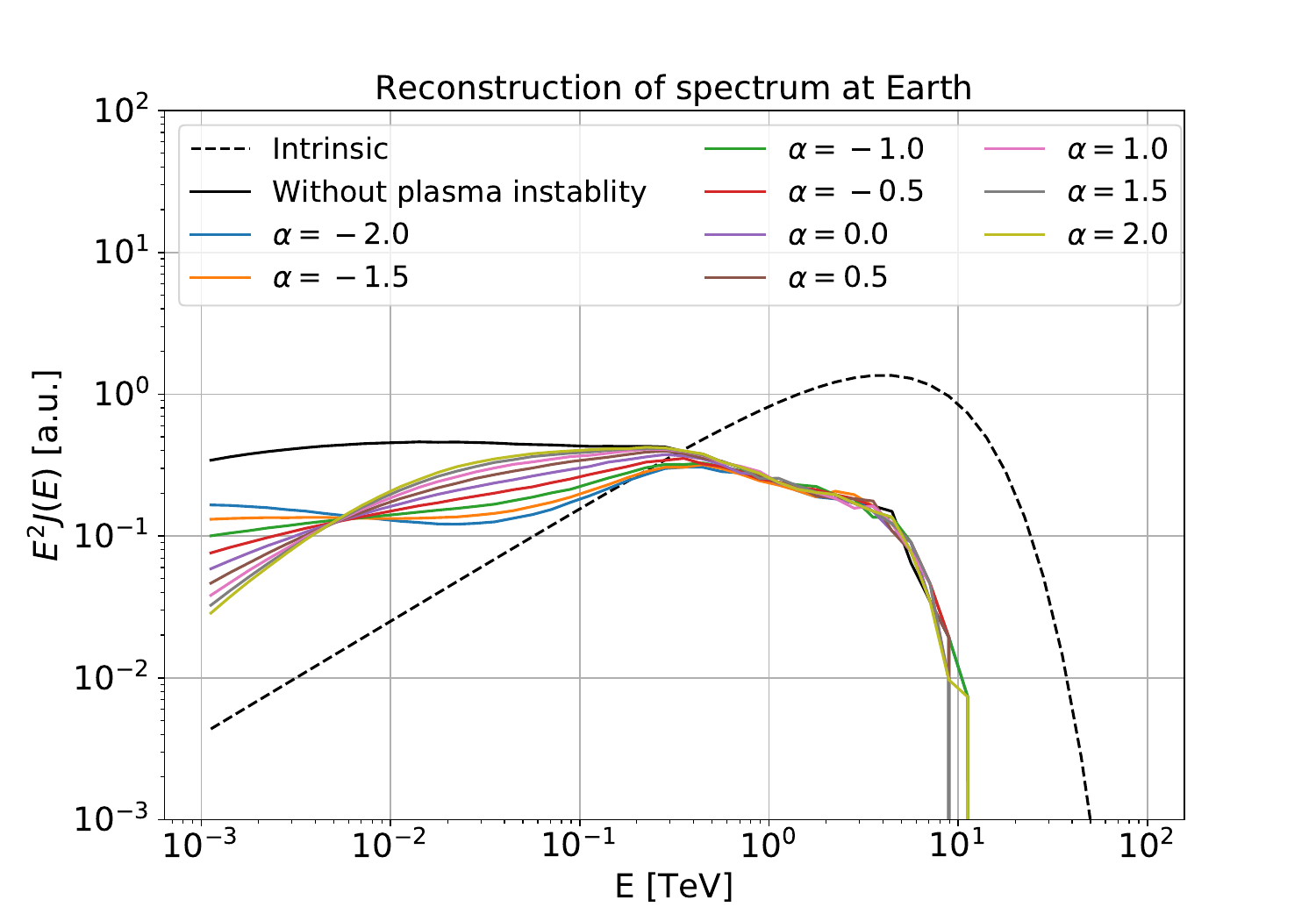}
\end{minipage}
\caption{Propagated photon energy spectra at Earth for an unattenuated spectrum (black dashed line) with spectral index $\beta=1.2$, exponential cutoff at 5.0 TeV (see \eqref{eq:3.3}). 
The plasma instability parameters are $\tilde{E} = 1.0$ TeV and $\lambda_0 = 0.12$ kpc (UPPER LEFT), $\lambda_0 = 1.2$ kpc (UPPER RIGHT), $\lambda_0 = 12.0$ kpc (BOTTOM LEFT) and $\lambda_0 = 120.0$ kpc (BOTTOM RIGHT). The coloured lines represent different values of the plasma index $\alpha$. Solid black line represents the energy spectrum in the scenario without plasma instabilities.}
\label{fig:fig3}
\end{figure}
We initially consider an intrinsic photon energy spectrum characterized by the parameters $\beta = 1.2$ and $E_{\text{cut}} = 5.0$ TeV. They corresponds to the best fit parameters for the blazar 1ES 0229+200 from \cite{AlvesBatista2019}. We compute the propagated photon spectrum for different combinations of instability power index $\alpha$ and length scale $\lambda_0$. The values of the plasma instability parameters used in this section are discussed in Section \ref{sec_model}.
\par In Figure \ref{fig:fig3} the photon spectra for all combinations of instability parameters are shown. The blazar unattenuated spectrum (dashed black line) and the propagated spectrum for the non-instability scenario (solid black line) are also shown as references. Different coloured lines correspond to different values of the plasma index $\alpha$. In the upper panels of Figure \ref{fig:fig3} ($\lambda_0=0.12\,\text{kpc}$ left and $\lambda_0=1.2\,\text{kpc}$ right) we see that all the values of the plasma instability index $\alpha$ produce similar results. A strong suppression of secondary photons in the GeV energy range is observed, suggesting a GeV-spectrum almost entirely composed of unattenuated photons. This can be understood by considering that instabilities dominate over IC, since their typical length scale is $\lesssim \lambda_{IC}$ in the energy range considered (see Figure \ref{fig:fig1}). The suppression shows the tendency of the propagated spectrum to approach the unattenuated spectrum from above. It is clearly not possible to suppress the photon spectrum below the unattenuated one, as this part is populated by photons that do not produce electron-positron pairs. 
\par In the bottom panels of Figure \ref{fig:fig3} the propagated spectra for $\lambda_0=12.0\,\text{kpc}$ on the left and $\lambda_0=120.0\,\text{kpc}$ on the right are shown. We can see that the propagated spectra start to deviate from the unattenuated emission at GeV scale. The observed spectra show a suppression for $E\lesssim100\,\text{GeV}$ with respect to the case without plasma instabilities. Two other main features can be observed.
Firstly, for $\alpha>0$, the suppression of the spectrum continues down to the GeV scale. This is due to the fact that for electron-positron pairs plasma instabilities are still more efficient than IC scattering at these energies. Secondly, a relative increase of the photon spectrum is observed for negative $\alpha$ at $E\sim10\,\text{GeV}$. In this case, the energy of the electron-positron pairs decreases to the point where IC again becomes the dominant process.
\par The inversion point between the different behaviours of the observed photon spectra described above occurs at $E\sim10\,\text{GeV}$. This is due to the relation between the energy of the up-scattered photon and the energy of the electron in the IC scattering (see \eqref{eq:2.7}). As discussed in Section \ref{sec_model}, for the interaction between 1.0 TeV electrons and CMB photons we expect up-scattered photons with energy $E\sim10\,\text{GeV}$. Electrons at energies around 1.0 TeV are also responsible for the transition of dominance of prompt photons above a few 100 GeV to cascade photons at lower energies, that can be seen in Figure \ref{fig:fig3}.
\par
\begin{figure}[t]
\centering
\begin{minipage}{7.5cm}
\centering
\includegraphics[scale=0.3]{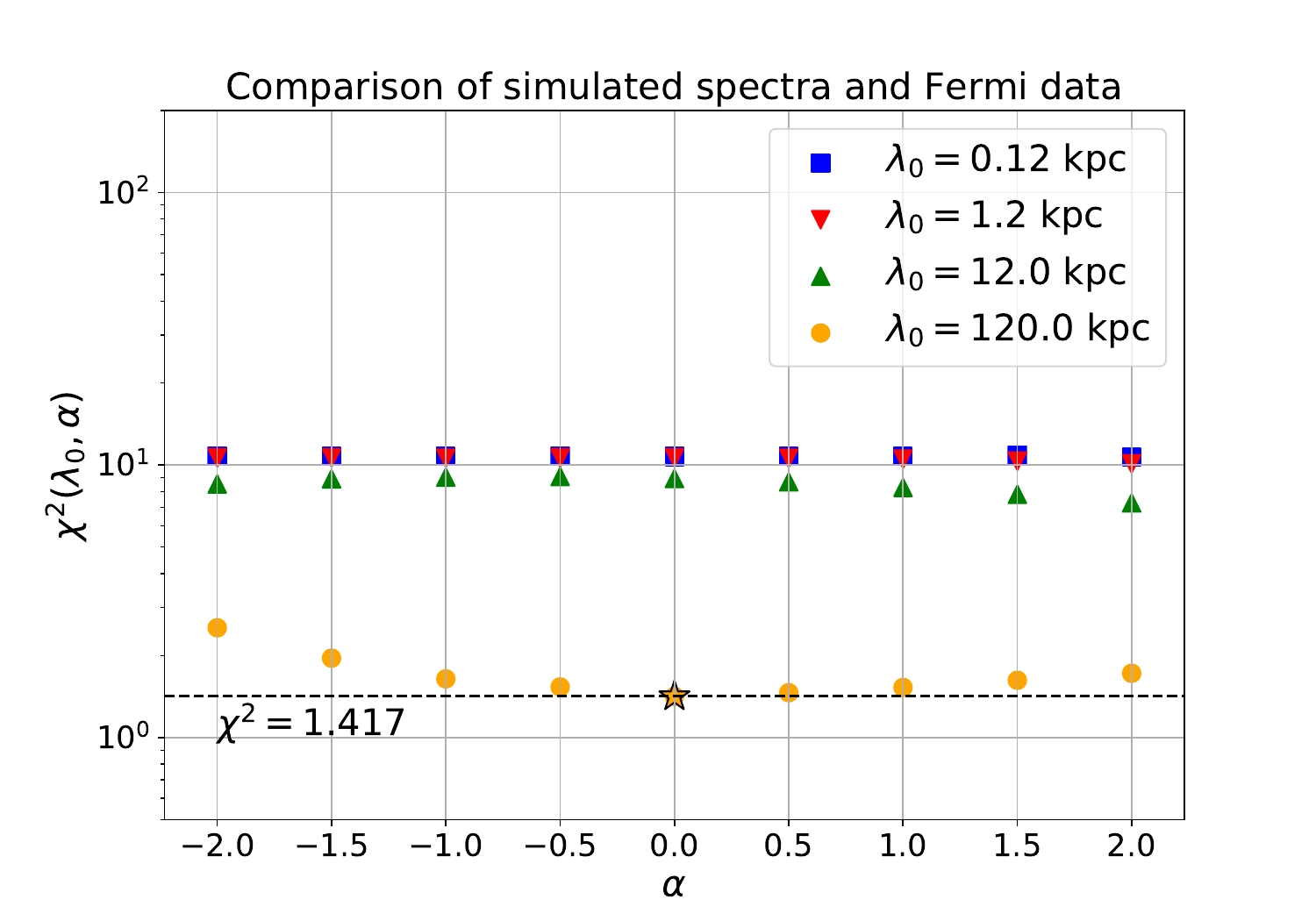}
\end{minipage}
\begin{minipage}{7.5cm}
\centering
\includegraphics[scale=0.3]{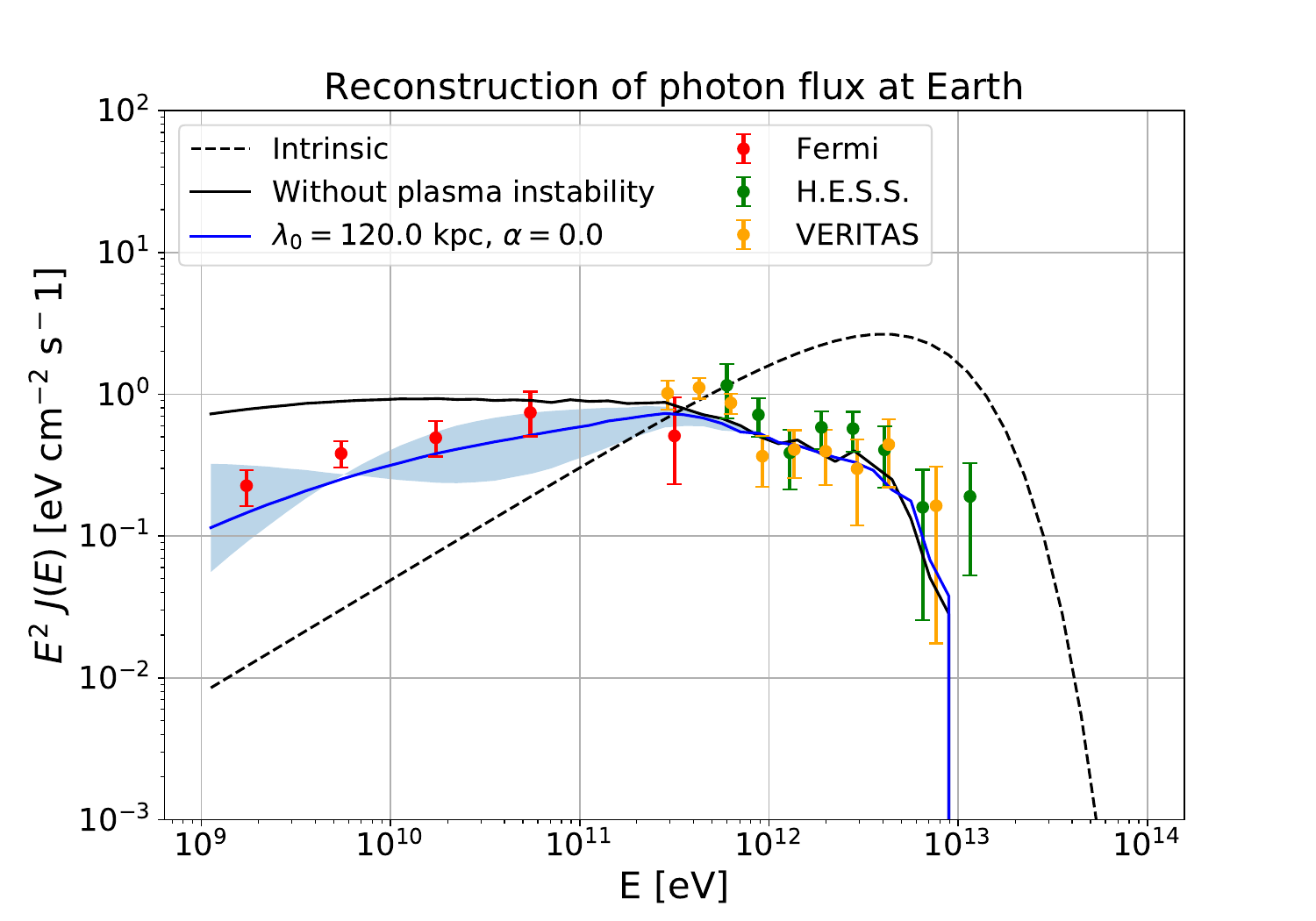}
\end{minipage}
\caption{LEFT: $\chi^2(\lambda_0,\alpha)$ values computed for Fermi data of the blazar 1ES 0229+200 and considered values of plasma instability parameters $\alpha$ and $\lambda_0$. The minimum value of $\chi^2$ is indicated with a yellow star and black dashed line. The minimum value is also indicated in the figure. RIGHT: unattenuated injected spectrum (black dashed line), propagated spectrum without instabilities (black solid line), propagated spectrum with instability parameters corresponding to the best agreement with the data (blue solid line) and Fermi sub-TeV data from \cite{Acciari2022} (red points). Blue shaded area corresponds to the propagated spectra for $\lambda_0=120.0\,\text{kpc}$ and all values of $\alpha$. TeV data from H.E.S.S. (form \cite{Aharonian2007}, green points) and VERITAS (from \cite{Aliu2014}, orange points) are also shown.}
\label{fig:figstat}
\end{figure}

 We compared our simulation results with the publicly available observed sub-TeV data by Fermi-LAT for the blazar 1ES 0229+200 within its 12-year source catalog\footnote{\href{https://fermi.gsfc.nasa.gov/cgi-bin/ssc/LAT/LATDataQuery.cgi}{https://fermi.gsfc.nasa.gov/cgi-bin/ssc/LAT/LATDataQuery.cgi}} (also available by the Spectral Energy Distribution Builder from the Space Science Data Center\footnote{\href{https://tools.ssdc.asi.it/SED/}{https://tools.ssdc.asi.it/SED/}}).  The propagated spectra are normalized with the best-fit  normalization value at the injection for the blazar to extract its photon flux. The corresponding unattenuated photon spectrum is then
\begin{equation}
\label{eq:4.1}
    J(E_0) = 1.94\cdot 10^{-24}\left(\frac{E_0}{1.0\,\text{TeV}}\right)^{-1.2}\exp\left(-\dfrac{E_0}{5.0\,\text{TeV}}\right)\,\text{eV}^{-1}\text{cm}^{-2}\text{s}^{-1}\, .
\end{equation}
We quantified the agreement between propagated and observed photon spectra by calculating the $\chi^2(\lambda_0,\alpha)$ for all the values of $\alpha$ and $\lambda_0$ considered in this work. We note that no TeV data were considered for the $\chi^2$ computation. The obtained $\chi^2$ values are shown in Figure \ref{fig:figstat} (left panel) and reported in Table \ref{tab:table} for all instability parameter combinations considered. We find the minimum $\chi^2\simeq1.4$ for the scenario with $\alpha=0.0$ and $\lambda_0=120.0$ kpc. These results show that plasma instabilities with length scales of the order of $\lambda_0\sim100\,\text{kpc}$ are able to suppress the photon flux in good agreement with observed data. We can also see that the instability power index $\alpha$ has a smaller effect on the $\chi^2$ than different length scales $\lambda_0$. In the right panel of Figure \ref{fig:figstat} we show the photon fluxes resulting from the $\chi^2$ study. In particular, the unattenuated spectrum (black dashed line), the propagated electromagnetic cascade (black solid line) and the spectrum obtained with the plasma parameters that minimize $\chi^2$ (blue solid line for the minimum and blue shaded area for $\lambda_0=120.0\,\text{kpc}$ and all values of $\alpha$) are shown along with observed data (red points). The TeV data for this blazar are also shown for completeness, taken by telescopes H.E.S.S. (from \cite{Aharonian2007}, green points) and VERITAS (from \cite{Aliu2014}, orange points). As already stated above, TeV data were not used for the $\chi^2$ analysis.
\begin{table}[t]
\centering
\begin{tabular}{|c|c|c|c|c|}
 \hline
 & $\lambda_0 = 0.12$ kpc & $\lambda_0 = 1.2$ kpc & $\lambda_0 = 12.0$ kpc & $\lambda_0 = 120.0$ kpc\\
 \hline
$\alpha=-2.0$  & 10.792 & 10.635 & 8.527 & 2.529 \\
$\alpha=-1.5$  & 10.781 & 10.623 & 8.907 & 1.958 \\
$\alpha=-1.0$  & 10.790 & 10.598 & 9.039 & 1.643 \\
$\alpha=-0.5$  & 10.764 & 10.629 & 9.079 & 1.535 \\
$\alpha=0.0$   & 10.747 & 10.637 & 8.925 & 1.417 \\
$\alpha=0.5$   & 10.784 & 10.594 & 8.682 & 1.462 \\
$\alpha=1.0$   & 10.794 & 10.503 & 8.286 & 1.528 \\
$\alpha=1.5$   & 10.831 & 10.333 & 7.813 & 1.623 \\
$\alpha=2.0$   & 10.700 & 10.103 & 7.267 & 1.722\\
 \hline
\end{tabular}
\caption{$\chi^2(\lambda_0,\alpha)$ values for propagated plasma instability scenarios.}
\label{tab:table} 
\end{table}
\par The effect of plasma instabilities is to reduce the electron energy before scattering with background photons. Since the energy of up-scattered photons depends on the energy of the in-coming electron (see \eqref{eq:2.7}), the amount of energy lost by electrons in plasma instabilities has a direct effect on the suppression of the propagated photon spectrum. We calculate the fraction of energy lost by the electrons in the scenario that best reproduces the observed photon flux (i.e. $\lambda_0=120.0$ kpc and $\alpha=0.0$). We integrate the energy-loss-per-unit-length of plasma instabilities in \eqref{eq:2.5} finding that the fraction of energy lost as a function of the propagated distance $x$ is given by\\
\begin{equation}
\label{eq:frac_ener}
    \left|\dfrac{E_e-E_{e,0}}{E_{e,0}}\right| = 1-\exp\left({-\dfrac{x-x_0}{\lambda_0}}\right)\, ,
\end{equation}
where $E_e(x=x_0)=E_{e,0}$ is the initial electron energy, $x_0$ is the position  where the electron is created, and $\lambda_0$ is the typical length scale of instabilities. The typical mean free path of IC scattering for GeV and TeV electrons is $\sim 1.2$ kpc. We obtain a fraction of energy lost over a distance of $\sim 1.2$ kpc equal to $\sim1\%$. This means that to reproduce the observed suppression of the photon flux, electrons and positrons must lose $\sim1\%$ of their energy in plasma instabilities within one IC scattering length. The total photon flux suppression will then be the result of this electron energy loss over many IC scattering lengths.
\par We highlight that the only blazar used for a direct comparison with observation data was 1ES 0229+4200. The observed GeV-photon spectra from other blazars, e.g. those considered in \cite{AlvesBatista2019}, are below what would be predicted by extrapolating at low energy the injected spectrum obtained from TeV data as a power law. We already discussed how plasma instabilities cannot reproduce such spectrum suppression. In these cases, the injected spectrum needs to fall below the extrapolated power law. This would introduce a degeneracy between injected and cascade spectrum, which does not allow to put constraints on cascade suppression by plasma instabilities.  Therefore, we do not consider these sources in our study.

\hypertarget{ResultsBlazars}{}
\subsection{Dependence on blazar injection parameters}
\label{subsec_results_blazars}
\begin{figure}[t]
\centering
\begin{minipage}{7.5cm}
\centering
\includegraphics[scale=0.3]{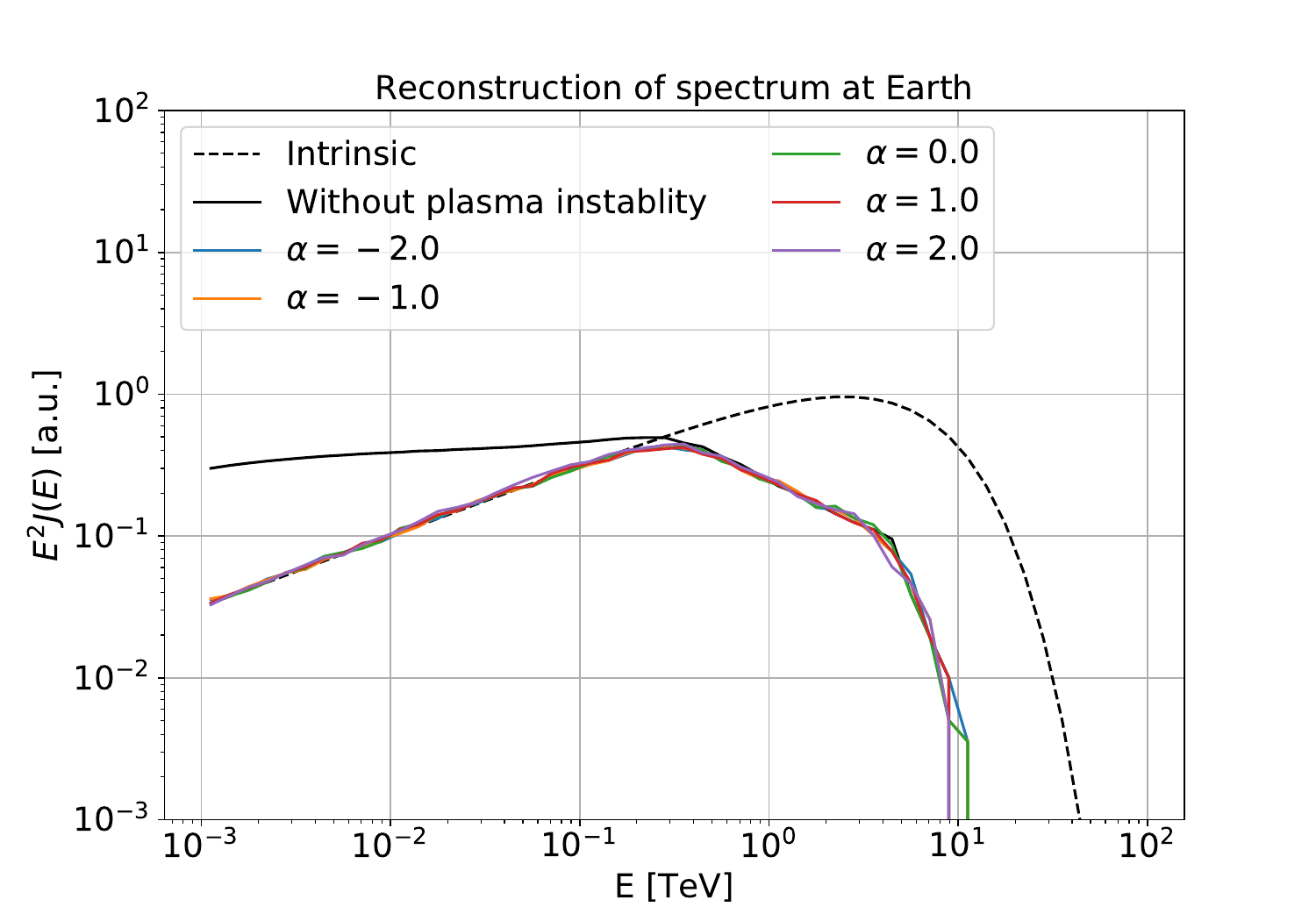}
\end{minipage}
\begin{minipage}{7.5cm}
\centering
\includegraphics[scale=0.3]{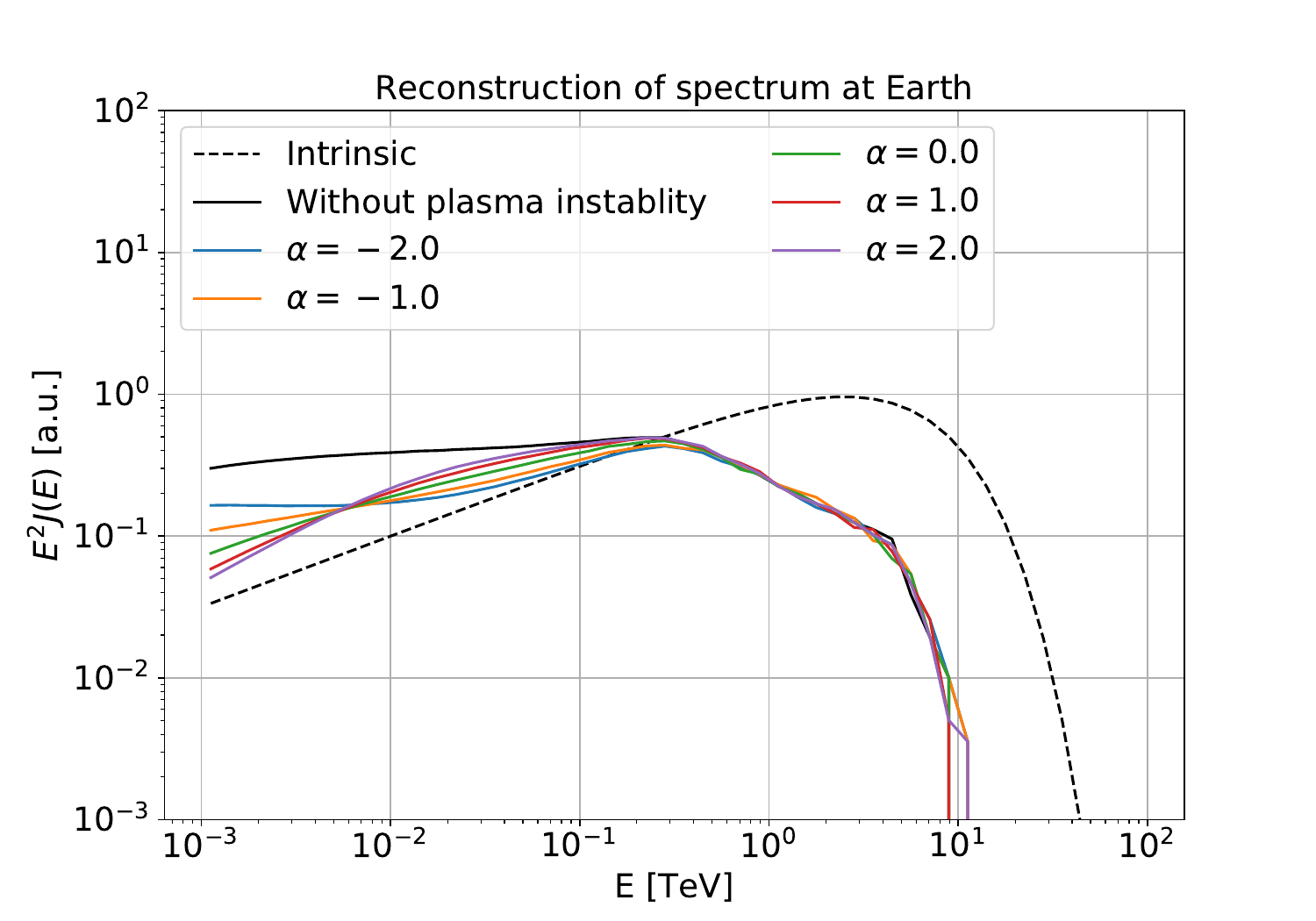}
\end{minipage}
\begin{minipage}{7.5cm}
\centering
\includegraphics[scale=0.3]{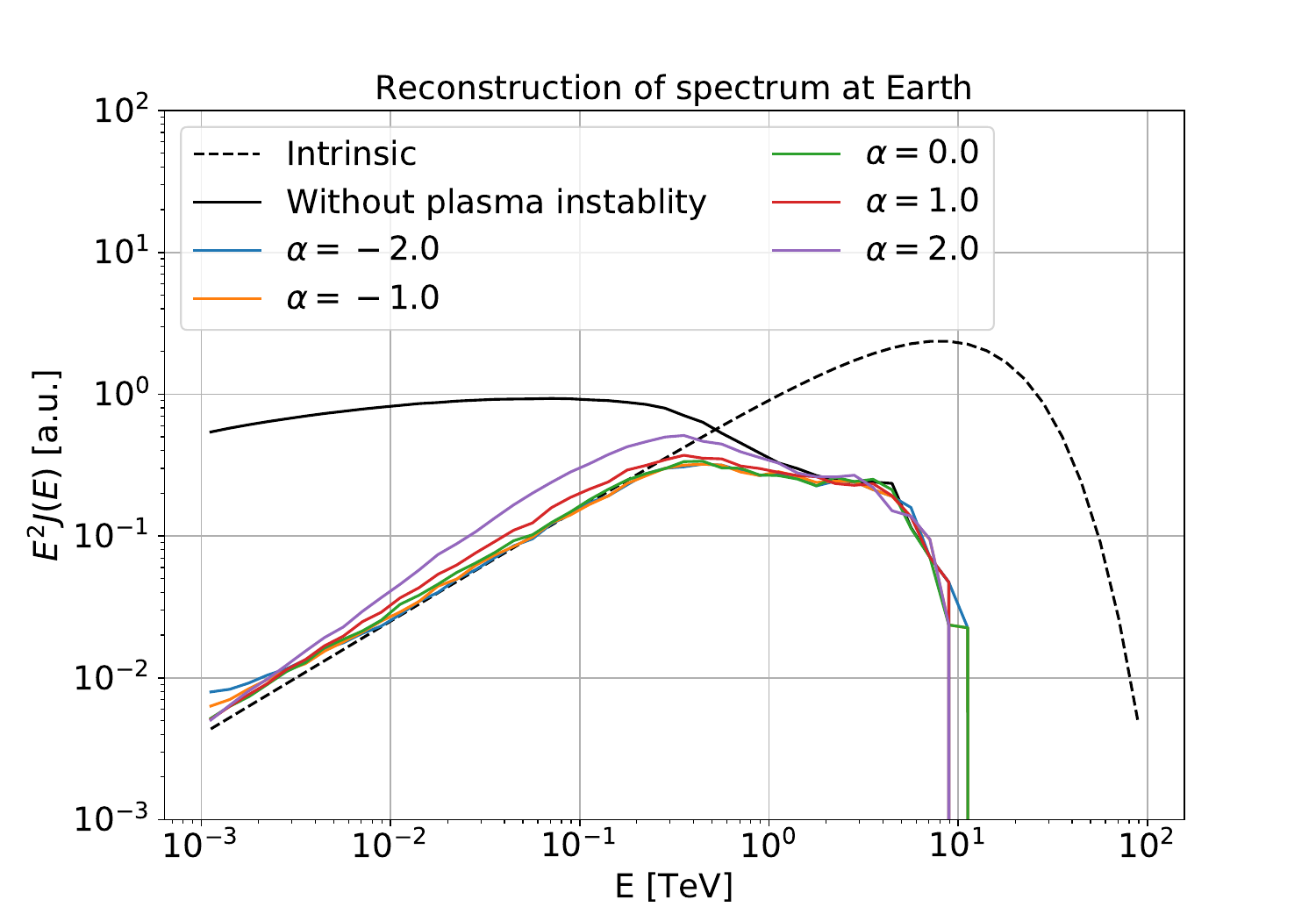}
\end{minipage}
\begin{minipage}{7.5cm}
\centering
\includegraphics[scale=0.3]{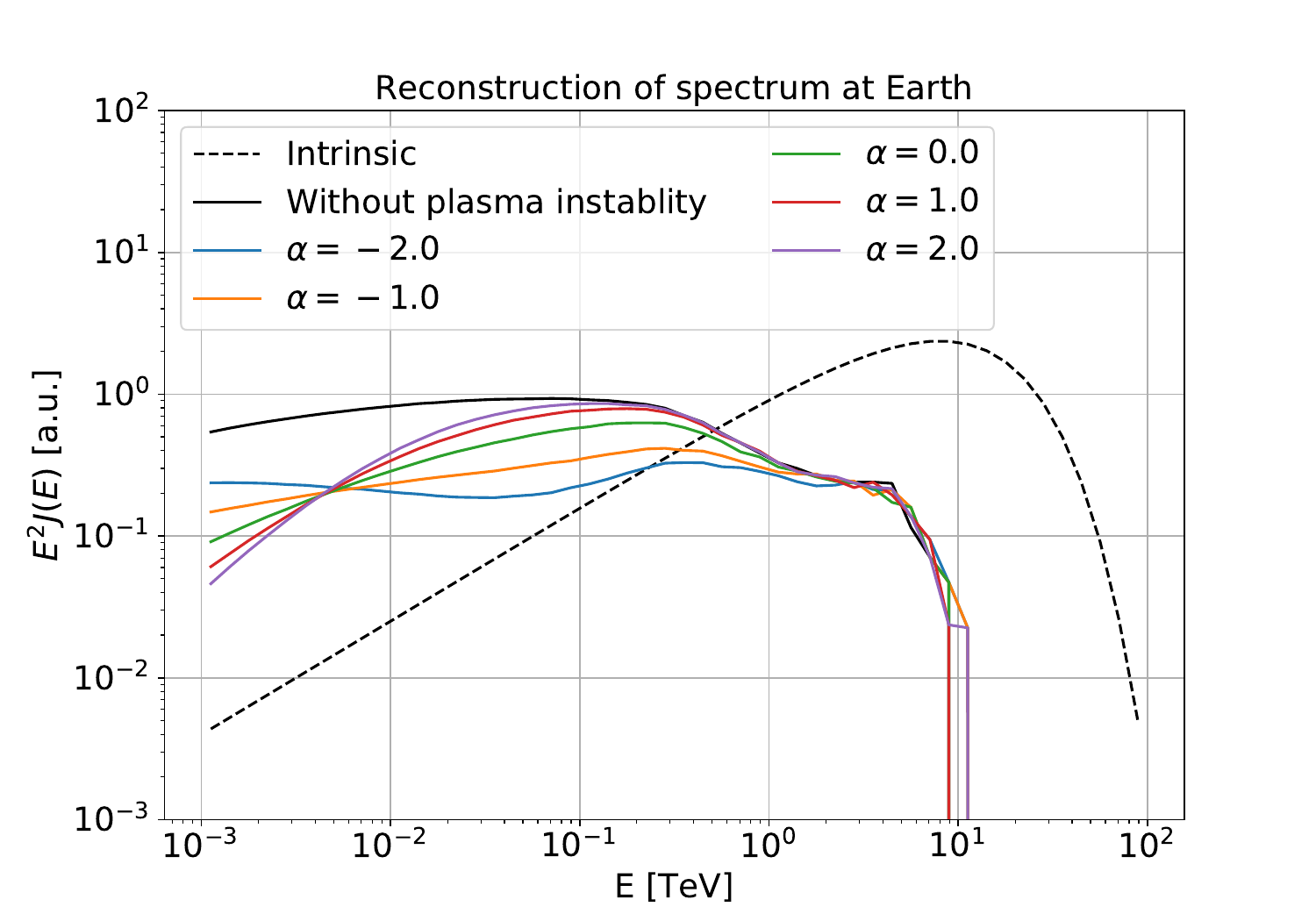}
\end{minipage}
\begin{minipage}{7.5cm}
\centering
\includegraphics[scale=0.3]{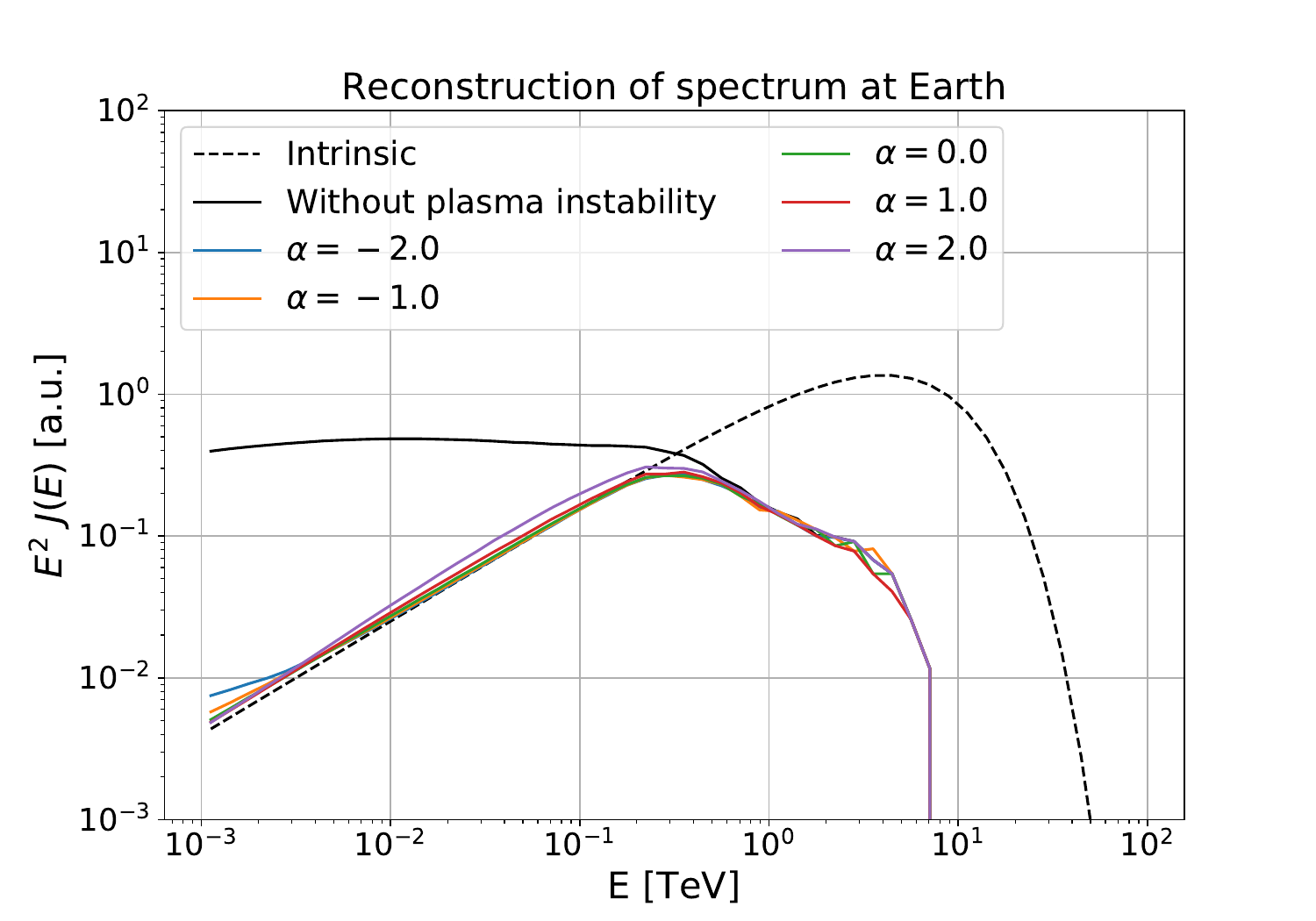}
\end{minipage}
\begin{minipage}{7.5cm}
\centering
\includegraphics[scale=0.3]{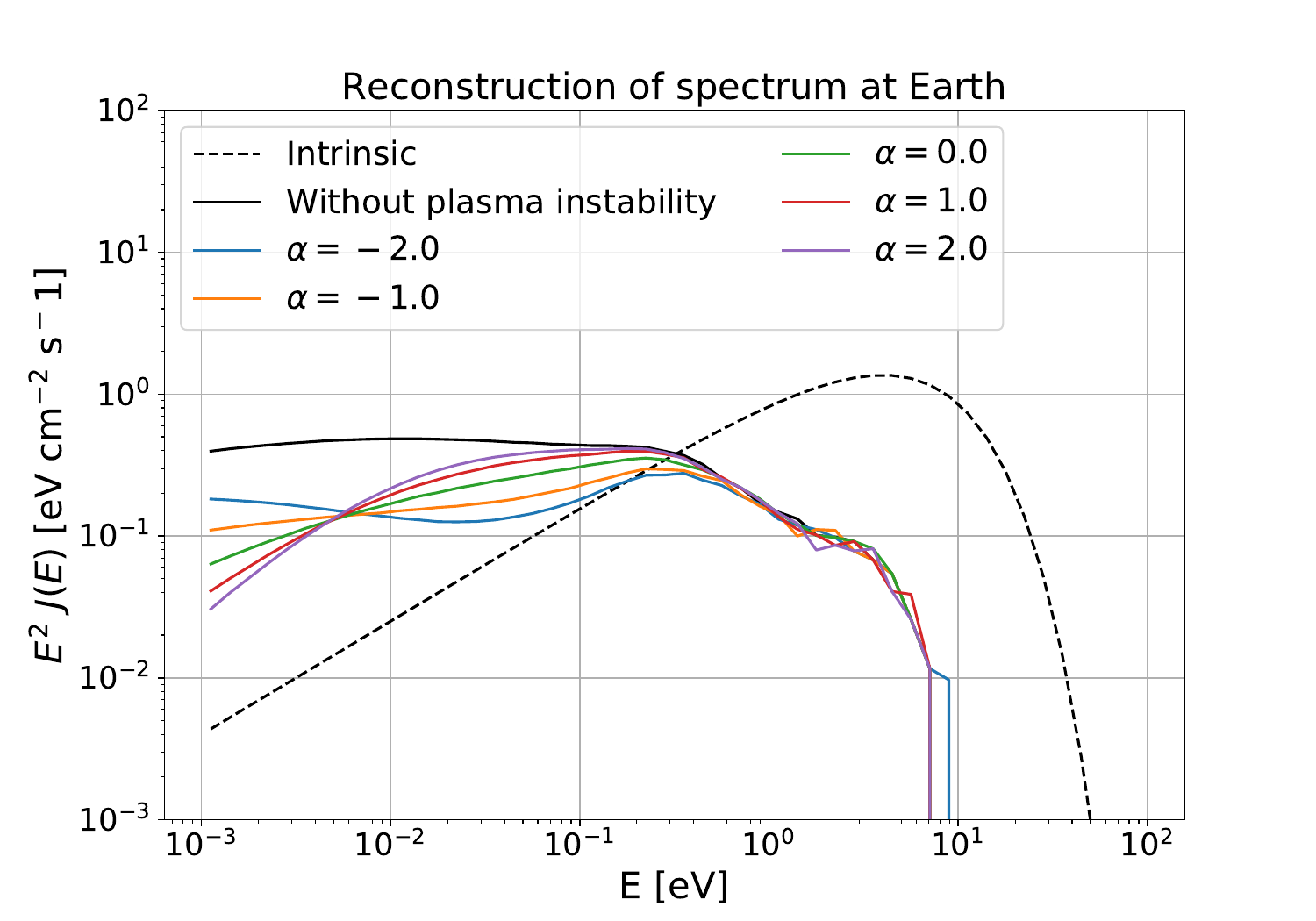}
\end{minipage}
\caption{Propagated photon spectra for different source parameterizaions (dashed black line): spectral index $\beta=1.5$, exponential cutoff at 5.0 TeV (UPPER), spectral index $\beta=1.2$, exponential cutoff at 10.0 TeV (CENTER) and spectral index $\beta=1.2$, exponential cutoff at 5.0 TeV (BOTTOM). The bottom panels correspond to the scenario with source distance of 800 Mpc. The parameters for the plasma instability energy loss are chosen as $\tilde{E} = 1.0$ TeV, $\lambda_0 = 1.2$ kpc (LEFT), $\lambda_0 = 120.0$ kpc (RIGHT). Coloured lines correspond to different values of the plasma index $\alpha$, while solid black line to the propagated spectrum in the scenario without instabilities.}
\label{fig:fig4}
\end{figure}
\par We continue our analysis by performing a parameter study of the blazar injection scenarios and simulating the corresponding electromagnetic cascades. In order to take into account strong variations in the results, we consider as source parameters the high energy cutoff $E_\text{cut}=10.0\,\text{TeV}$ and spectral index $\beta=1.5$ . In addition, we consider a new scenario where the source distance is $800\,\text{Mpc}$ (as one of the more distant blazar considered in \cite{AlvesBatista2019}). A new set of simulations is performed with the same configuration described in Section \ref{sec_simulations}, but with a observer radius of $800\,\text{Mpc}$. The computation of the photon spectrum at Earth follows the same procedure of Section \ref{subsec_output_analysis}. We evaluate the reconstructed photon spectra for $\lambda_0=1.2,$ kpc and $\lambda_0=120.0\,\text{kpc}$, and all the values of instability index $\alpha$. As before, only one parameter is varied at a time, keeping the others at their reference value \eqref{eq:4.1}.
\par We show the resulting propagated photon spectra in Figure \ref{fig:fig4}. As before, the blazar unattenuated spectrum (dashed black line) and the propagated spectrum for the non-instability scenario (solid black line) are shown for reference. Differently coloured lines correspond to different values of the plasma index $\alpha$. In the upper panels of Figure \ref{fig:fig4}, results for different injection spectral index $\beta$ are shown. We observe that the spectra at Earth for $\beta=1.5$ show a similarly strong suppression as for the case with $\beta=1.2$, when $\lambda_0=1.2$ kpc. In this case, for $\lambda_0=120.0$ kpc the resulting propagated spectra start to deviate from the non-instability scenario, but the observed suppression is now less than the one observed in Figure \ref{fig:fig3}. As mentioned in Section \ref{sec_intro}, the reason is that propagated electromagnetic cascades are characterized by an average spectral index of approximately 1.9 \cite{Berezinsky2016}. Therefore, sources with softer injection indices will be dominated by the prompt GeV photon flux. Only sources with spectral index $\beta < 1.9$ are suitable for investigating the influence of energy drain by plasma instabilities, since the GeV photon flux is dominated by the cascade particles.
\par In the central panels of Figure \ref{fig:fig4}, the value $E_{\text{cut}}=10.0\,\text{TeV}$ is used for the high energy cutoff. We observe that, when the value of $E_{\text{cut}}$ increases, the effect of the instabilities on the spectrum leads to slightly stronger dependence on the instability power index $\alpha$, beginning to be evident already for $\lambda_0=1.2\,\text{kpc}$. The explanation is that a higher value of $E_{\text{cut}}$ implies more TeV photons producing electron-positron pairs, thus initiating the cascade. In turn, a greater number of pairs subject to the influence of instabilities amplifies the differences between each instability configuration.
\par Finally, in the bottom panels of Figure \ref{fig:fig4}, the scenario for a source distance of 800 Mpc is shown. We do not observe significant changes to the overall shape of the photon spectrum at Earth with respect to the case in Figure \ref{fig:fig3} (source distance of $580\,\text{Mpc}$). This is consistent with the universality of the photon spectrum found in \cite{Berezinsky2016}: the spectral index of the cascade photon spectrum is independent on the distance to source when this is well above the Mpc-length scale. In comparison with Figure \ref{fig:fig3}, the spectra of the scenarios with and without plasma instabilities appear similar for both simulated propagation distances. Therefore, if the effect of the propagation distance is negligible, sources above a few hundreds of Mpc are suitable for the study of the energy-losses due to instabilities.
\par  Even for the examples of different blazar emission configurations considered in this section, the influence of plasma instabilities on the photon spectra at $E\sim10.0\,\text{GeV}$ is similar to that observed in Section \ref{subsec_results_parametric}. This confirms that this is not a feature of the source emission, but a consequence of the IC kinematics at the energy scale $\widetilde{E}_e=1.0,\text{TeV}$, considered in this work.

\hypertarget{Discussion and conclusions}{}
\section{Discussion and conclusions}
\label{sec_discussion}
In this work, we have studied the impact of the resonant interactions of electron-positron pairs with the IGM (i.e. plasma instabilities) during the development of electromagnetic cascades, initiated by the emission of TeV-photons from distant blazars. We modelled the energy loss term associated with plasma instabilities in Equation \eqref{eq:2.5}, and implemented it in the framework of the Monte Carlo code CRPropa 3.2. We performed a parametric study of the instability energy loss term by simulating the extragalactic propagation of high energy photons from the blazar 1ES 0229+200. In particular, we studied the induced suppression in the GeV photon spectrum by varying the plasma index $\alpha$ and typical length scale $\lambda_0$. Subsequently, the spectrum parameters of the injected photons were also varied to study their impact on the observable suppression of the spectrum. 
\par Results from blazar 1ES 0229+200 in Figure \ref{fig:fig3} have shown that instabilities can induce a non-negligible suppression in the GeV energy range of the propagated photon spectrum, for different values of the instability length scale $\lambda_0$. A natural upper bound for the instability length scale, in order to produce an observable effect, corresponds to the distance of the source $L$ (which for the case discussed here was of the order $500$ Mpc). This can be clearly seen in Figure \ref{fig:fig3}, where the photon spectrum becomes less and less suppressed (compared to the case without instabilities) as the instability length scale $\lambda_0$ converges to $L$. However, numerical simulations have shown that the suppression may also become negligible for length scales which are some orders of magnitude below $L$. The reason is that the typical length scale of instabilities is such that the IC scattering completely dominates. 
\par Two groups of plasma instabilities can be identified based on their energy dependence: those with $\alpha > 0$ or $\alpha < 0$. These two groups of instabilities show different trends in the photon energy spectra at Earth. Models whose energy loss length increases at higher energies ($\alpha>0$) start to resemble the cascade spectrum for energies above $\sim10.0\,\text{GeV}$. On the contrary, models whose energy loss length decreases at higher energies ($\alpha<0$) show a stronger suppression above $\sim10.0\,\text{GeV}$. However, below the inversion point, at around $10.0\,\text{GeV}$, the two groups of instabilities begin to show opposite behaviour: for $\alpha>0$ the suppression continues down to $\sim1.0\,\text{GeV}$, while the photon spectrum in the models with $\alpha<0$ begins to increase, converging to the cascade spectrum. The sign and absolute value of the plasma index $\alpha$ has a crucial role for the determination of the type of instabilities developed in the IGM. For instance, for large and positive values of the plasma index ($\alpha > 1$), models tend to consider a dominance of the modulation instability for the energy losses, as has been previously studied in \cite{Schlickeiser2012,Vafin2018}. On the contrary, most models which consider the instabilities to be governed by oblique instabilities report small or negative spectral indices, such as in \cite{Broderick2012,Sironi2014}.
\par In Section \ref{subsec_results_blazars}, we have shown that the intrinsic blazar emission parameters (in particular the spectral index $\beta$) are essential factors for studying the effect of different plasma models on the propagated cascade. In general, only blazars whose spectrum is not dominated by prompt photons at GeV energies are suitable for probing plasma instabilities. This leads to the requirement $\beta< 1.9$, with 1.9 being the characteristic spectral index for fully developed electromagnetic cascades from primary photons subject to pair production.
\par A comparative analysis with Fermi-LAT data of 1ES 0229+200 showed that for $\lambda_0\sim\mathcal{O}(100$ kpc), plasma instabilities suppress the photon spectrum  below $\sim100\,$GeV sufficiently to reproduce the observed flux. In particular, the simulated spectra at Earth for $\lambda_0=120$ kpc and $\alpha=0$ show the best agreement with the gamma-ray data out of all considered parameter combinations. The fraction of electron energy lost over the typical IC interaction length, for this set of parameters, was found to be $\sim1\%$. This value corresponds to the upper limit of recent theoretical estimates, such as those in \cite{Alawashra2024}, which suggest that the maximum fraction of energy lost is at the percent level or smaller. Furthermore, from Figure \ref{fig:figstat} we can see that different values of $\alpha$ still show good agreement with the data if $\lambda_0$ does not significantly change. Therefore, we can conclude that plasma instabilities with energy-loss lengths of the order of $\sim100$ kpc may have a relevant role in the suppression of the energy spectrum of TeV-blazars for energies in the GeV range.
\par Similar studies combining plasma instabilities and magnetic deflections could help to impose more accurate constrains on the IGMF. Blazars with a hard spectral index, such as 1ES 0229+200, are expected to be good candidates for these studies since their propagated spectra is dominated by cascade photons. Three-dimensional simulations are required to take into account the IGMF and different scenario for  momentum distribution at the injection. CRPropa offers several tools to use synthetic IGMF and explore different injection scenarios (e.g. directed emission following a von-Mises-Fisher distribution, see \cite{AlvesBatista2022,Jasche2019}). Future works, including the IGMF, may shed light on the possibilities of untangling these two different phenomena. Therefore, estimates of the instability power index are required for more conclusive studies. Moreover, instability-induced angular spreading should also be included in simulations. This effect will allow to compare the magnetic and instability widening of the cascade, as well as their possible interplay. Studies including these terms will be of fundamental importance to examine the suppression in a time-dependent framework, and compare it with the IGMF suppression due to image broadening and time delays of secondary photons.

\acknowledgments
We acknowledge support by the Deutsche Forschungsgemeinschaft (DFG, German Research Foundation) under Germany’s Excellence Strategy -- EXC 2121 ``Quantum Universe'' -- 390833306,
and by the Bundesministerium für Bildung und Forschung, under grant 05A20GU2.


\begin{thebibliography}{99}
\bibliographystyle{abbrv}

\bibitem{Dicke1965} \hypertarget{Dicke1965}{R. H. Dicke, P. J. E. Peebles, P. G. Roll and D. T. Wilkinson, \textit{Cosmic Black-Body Radiation.} ApJ, \textbf{142} (1965) 414 [\href{https://ui.adsabs.harvard.edu/abs/1965ApJ...142..414D/abstract}{10.1086/148306}]}

\bibitem{Penzias1965} \hypertarget{Penzias1965}{A. A. Penzias and R. W. Wilson, \textit{A Measurement of Excess Antenna Temperature at 4080 Mc/s.} ApJ, \textbf{142} (1965) 419 [\href{https://ui.adsabs.harvard.edu/abs/1965ApJ...142..419P/abstract}{ 
10.1086/14830}]}

\bibitem{Hauser2001} \hypertarget{Hauser2001}{M. G. Hauser and E. Dwek, \textit{The Cosmic Infrared Background: Measurements and Implications.} ARA$\&$A, \textbf{39} (2001) 249 [\href{https://arxiv.org/abs/astro-ph/0105539}{astro-ph/0105539}]}

\bibitem{Heiter:2017cev} \hypertarget{Heiter:2017cev}
{C.~Heiter, D.~Kuempel, D.~Walz and M.~Erdmann, \textit{Production and propagation of ultra-high energy photons using CRPropa 3.} Astropart. Phys. \textbf{102} (2018), 39-50 [\href{https://arxiv.org/pdf/1710.11406}{astro-ph/1710.11406}]}

\bibitem{Berezinsky2016} \hypertarget{Berezinsky2016}{V. Berezinsky and O. Kalashev, \textit{High energy electromagnetic cascades in extragalactic space: physics and features.} Phys. Rev. D, \textbf{94} (2016) 023007 [\href{https://arxiv.org/abs/1603.03989}{astro-ph/1603.03989}]}

\bibitem{Krawczynski1999} \hypertarget{Krawczynski1999}{W. Hofmann and Collaboration H.E.S.S. Collaboration, \textit{The high energy stereoscopic system (HESS) project.} AIP Conf. Proc., \textbf{515} (2000) 500 [\href{https://pubs.aip.org/aip/acp/article/515/1/500/579616/The-high-energy-stereoscopic-system-HESS-project}{10.1063/1.1291416}]}

\bibitem{Saggion2006} \hypertarget{Saggion2006}{J. Rico, \textit{Review of fundamental physics results with the MAGIC telescopes.} AIP Conf. Proc, \textbf{1792} (2017) 060001 [\href{https://pubs.aip.org/aip/acp/article/1792/1/060001/885991/Review-of-fundamental-physics-results-with-the}{10.1063/1.4968984}]}

\bibitem{Holder2006} \hypertarget{Holder2006}{T. C. Weekes et al., \textit{VERITAS: the Very Energetic Radiation Imaging Telescope Array Systeme.} Astropart. Phys., \textbf{17} (2002) 221 [\href{https://arxiv.org/abs/astro-ph/0108478}{astro-ph/0108478}]}

\bibitem{Atwood2009} \hypertarget{Atwood2009}{W. B. Atwood et al., \textit{The Large Area Telescope on the Fermi Gamma-ray Space Telescope Mission.} ApJ, \textbf{697} (2009) 1071 [\href{https://arxiv.org/abs/0902.1089}{astro-ph/10902.1089}]}

\bibitem{Acciari2022} \hypertarget{Acciari2022}{MAGIC Collaboration, A. Neronov, D. Smeikoz and A. Korochkin, \textit{A lower bound on intergalactic magnetic fields from time variability of 1ES 0229+200 from MAGIC and Fermi/LAT observations.} A$\&$A, \textbf{670} (2022) A145 [\href{https://arxiv.org/abs/2210.03321}{astro-ph/2210.03321}]}

\bibitem{Vovk2023} \hypertarget{Vovk2023}{I. Vovk, A. Korochkin, A. Neronov, D. Semikoz, \textit{Constraint on intergalactic magnetic field from Fermi/LAT observations of the "pair echo" of GRB 221009A.} (2023) [\href{https://arxiv.org/abs/2306.07672}{astro-ph/2306.07672}]}

\bibitem{DaVela2023} \hypertarget{DaVela2023}{P. Da Vela, G. Martí-Devesa, F. G. Saturni, P. Veres, A. Stamerra and F. Longo, \textit{Intergalactic magnetic field studies by means of $\gamma$-ray emission from GRB 190114C.} Phys. Rev. D, \textbf{107} (2023) 063030 [\href{https://arxiv.org/abs/2303.03137}{astro-ph/2303.03137}]}

\bibitem{Korochkin2021} \hypertarget{Korochkin2021}{A. Korochkin, O. Kalashev, A. Neronov and D. Semikoz, \textit{Sensitivity reach of gamma-ray measurements for strong cosmological magnetic fields.} ApJ., \textbf{906} (2021) 116 [\href{https://arxiv.org/abs/2007.14331}{astro-ph/2007.14331}]}

\bibitem{Neronov2009} \hypertarget{Neronov2009}{A. Neronov and D. Semikoz, \textit{Sensitivity of gamma-ray telescopes for detection of magnetic fields in intergalactic medium.} Phys. Rev. D, \textbf{80} (2009) 123012 [\href{https://arxiv.org/abs/0910.1920}{astro-ph/0910.1920}]}

\bibitem{Neronov2010} \hypertarget{Neronov2010}{A. Neronov and I. Vovk, \textit{Evidence for strong extragalactic magnetic fields from Fermi observations of TeV blazars.} Sci., \textbf{328} (2010) 73 [\href{https://arxiv.org/abs/1006.3504}{astro-ph/1006.3504}]}

\bibitem{Taylor2011} \hypertarget{Taylor2011}{A. M. Taylor, I. Vovk and A. Neronov, \textit{EGMF Constraints from Simultaneous GeV-TeV Observations of Blazars.} A$\&$A, \textbf{529} (2011) A114 [\href{https://arxiv.org/abs/1101.0932}{astro-ph/1101.0932}]}

\bibitem{Tonks1929} \hypertarget{Tonks1929}{L. Tonks and I. Langmuir, \textit{Oscillations in Ionized Gases.} Phys. Rev., \textbf{33} (1929) 195 [\href{https://journals.aps.org/pr/abstract/10.1103/PhysRev.33.195}{10.1103}]}

\bibitem{Alawashra2024} \hypertarget{Alawashra2024}{M. Alawashra and M. Pohl, \textit{Nonlinear feedback of the electrostatic instability on the blazar-induced pair beam and GeV cascade.} (2024) [\href{https://arxiv.org/abs/2402.03127}{astro-ph/2402.03127}]}

\bibitem{Perry2021} \hypertarget{Perry2021}{R. Perry, Y. Lyubarsky, \textit{The role of resonant plasma instabilities in the evolution of blazar induced pair beams.} MNRAS, \textbf{503} (2021) 2215 [\href{https://arxiv.org/abs/2102.03190}{astro-ph/2102.03190}]}

\bibitem{Alawashra2022} \hypertarget{Alawashra2022}{M. Alawashra and M. Pohl, \textit{Suppression of the TeV pair-beam plasma instability by a tangled weak intergalactic magnetic field.} ApJ, \textbf{929} (2022) 67 [\href{https://arxiv.org/abs/2203.01022}{astro-ph/2203.01022}]}

\bibitem{AlvesBatista2019} \hypertarget{AlvesBatista2019}{R. Alves Batista, A. Saveliev and E. M. de Gouveia Dal Pino, \textit{The Impact of Plasma Instabilities on the Spectra of TeV Blazars.} MNRAS, \textbf{489}, (2019) 3836 [\href{https://arxiv.org/abs/1904.13345}{astro-ph/1904.13345}]}

\bibitem{AlvesBatista2016} \hypertarget{AlvesBatista2016}
{R. Alves Batista et al., \textit{CRPropa 3 - a Public Astrophysical Simulation Framework for Propagating Extraterrestrial Ultra-High Energy Particles.} JCAP \textbf{05} (2016), 038 [\href{https://arxiv.org/abs/1603.07142}{astro-ph/1603.07142}]} 

\bibitem{AlvesBatista2022} \hypertarget{AlvesBatista2022}{R. Alves Batista et al., \textit{CRPropa 3.2 -- an advanced framework for high-energy particle propagation in extragalactic and galactic spaces.} JCAP, \textbf{2209} (2022) 035 [\href{https://arxiv.org/abs/2208.00107}{astro-ph/2208.00107}]} 

\bibitem{Nicholson1983} \hypertarget{Nicholson1983}{D. R. Nicholson, \textit{Introduction to Plasma Theory.} JWS (1983) 60-69}

\bibitem{Beck2023} \hypertarget{Beck2023}{M. Beck, O. Ghosh, F. Grüner, M. Pohl, C. B. Schroeder, G. Sigl, R. D. Stark and B. Zeitler, \textit{Evolution of Relativistic Pair Beams: Implications for Laboratory and TeV Astrophysics.} (2023) [\href{https://arxiv.org/abs/2306.16839}{astro-ph/2306.16839}]}

\bibitem{Franceschini2008} \hypertarget{Franceschini2008}{A. Franceschini, G. Rodighiero and M. Vaccari, \textit{The extragalactic optical-infrared background radiations, their time evolution and the cosmic photon-photon opacity.} A$\&$A, \textbf{487} (2008) 837 [\href{https://arxiv.org/abs/0805.1841}{astro-ph/0805.1841}]}

\bibitem{Jackson1998}\hypertarget{Jackson1998}{J. D. Jackson, \textit{Classical Electrodynamics}, JohnWiley \& Sons (1998)}

\bibitem{Sigl2017} \hypertarget{Sigl2017}{G. Sigl, \textit{Astroparticle Physics: Theory and Phenomenology.} ATLANTISSAP, \textbf{1} (2017)}

\bibitem{Fixsen2009} \hypertarget{Fixsen2009}{D. J. Fixsen, \textit{The Temperature of the Cosmic Microwave Background.} ApJ, \textbf{707} (2009) 916 [\href{https://arxiv.org/abs/0911.1955}{astro-ph/0911.1955}]}

\bibitem{Aharonian2007} \hypertarget{Aharonian2007}{F. Aharonian et al., \textit{New constraints on the Mid-IR EBL from the HESS discovery of VHE gamma rays from 1ES 0229+200.} A$\&$A, \textbf{475} (2007) 2 [\href{https://arxiv.org/abs/0709.4584}{astro-ph/0709.4584}]}

\bibitem{Aliu2014} \hypertarget{Aliu2014}{E. Aliu et al., \textit{A Three-Year Multi-Wavelength Study of the Very High Energy Gamma-ray Blazar 1ES 0229+200.} ApJ, \textbf{782} (2014) 13 [\href{https://arxiv.org/abs/1312.6592}{astro-ph/1312.6592}]}

\bibitem{Schlickeiser2012} \hypertarget{Schlickeiser2012}{R. Schlickeiser, D. Ibscher and M. Supsar, \textit{Plasma Effects on Fast Pair Beams in Cosmic Voids.} ApJ, \textbf{758} (2012) 102 [\href{https://iopscience.iop.org/article/10.1088/0004-637X/758/2/102}{ 
10.1088/0004-637X/758/2/102}]}

\bibitem{Vafin2018} \hypertarget{Vafin2018}{S. Vafin, I. Rafighi, M. Pohl, and J. Niemiec, \textit{The electrostatic instability for realistic pair distributions in blazar/EBL cascades.} ApJ, \textbf{857} (2018) 43 [\href{https://arxiv.org/abs/1803.02990}{astro-ph/1803.02990}]}

\bibitem{Broderick2012} \hypertarget{Broderick2012}{A. E. Broderick, P. Chang and C. Pfrommer, \textit{The Cosmological Impact of Luminous TeV Blazars I: Implications of Plasma Instabilities for the Intergalactic Magnetic Field and Extragalactic Gamma-Ray Background.} ApJ, \textbf{752} (2012) 22 [\href{https://arxiv.org/abs/1106.5494}{astro-ph/1106.5494}]}

\bibitem{Sironi2014} \hypertarget{Sironi2014}{L. Sironi and D. Giannios, \textit{Relativistic Pair Beams from TeV Blazars: A Source of Reprocessed GeV Emission rather than IGM Heating.} ApJ, \textbf{787} (2014) 49 [\href{https://arxiv.org/abs/1312.4538}{astro-ph/1312.4538}]}

\bibitem{Jasche2019} \hypertarget{Jasche2019}{J. Jasche, A. van Vliet and J. P. Rachen, \textit{Targeting Earth: CRPropa learns to aim.} PoS, \textbf{358} (2019) [\href{https://arxiv.org/abs/1911.05048}{astro-ph/1911.05048}]}

\end{thebibliography}
\end{document}